\documentclass[journal]{IEEEtran}

\usepackage{amsmath,amssymb,amsfonts,amsthm}
\usepackage{bm}
\usepackage{cite}
\usepackage{graphicx}
\usepackage{booktabs}
\usepackage{algorithm}
\usepackage{algorithmic}
\usepackage{subfigure}
\usepackage{xcolor}

\usepackage[
  colorlinks=true,
  linkcolor=blue,
  citecolor=blue,
  urlcolor=blue
]{hyperref}

\newcommand{\mat}[1]{\mathbf{#1}}
\newcommand{\vect}[1]{\mathbf{#1}}
\newcommand{\FM}{\mat F_{\mathrm M}}
\newcommand{\Range}{\operatorname{Range}}

\newcommand{\rank}{\operatorname{rank}}
\newcommand{\argmin}{\operatornamewithlimits{arg\,min}}

\newcommand{\CN}{\mathcal{CN}}
\newcommand{\algref}[1]{\textbf{Algorithm~\ref{#1}}}
\newcommand{\propref}[1]{\textbf{Proposition~\ref{#1}}}
\newcommand{\lemref}[1]{\textbf{Lemma~\ref{#1}}}
\newcommand{\figref}[1]{Fig.~\ref{#1}}

\newtheoremstyle{ieeeplain}{1pt}{1pt}{\itshape}{\parindent}{\bfseries}{.}{5pt}{}
\theoremstyle{ieeeplain}
\newtheorem{lemma}{Lemma}
\newtheorem{proposition}{Proposition}

\newtheorem{remark}{Remark}

\begin{document}

\title{Symbol-Level Precoding With Microwave Linear Analog Computers (MiLACs): Performance Limits and Stem-Connected Hardware Design}

\author{Yafei Wang, Yuchen Zhang, \textit{Member, IEEE},
Yiming Zhu, Jiayang Wan, Wenjin Wang, \textit{Member, IEEE},\\
Symeon Chatzinotas, \textit{Fellow}, \textit{IEEE}, and Björn Ottersten, \textit{Fellow}, \textit{IEEE}
\thanks{Yafei Wang, Yiming Zhu, Jiayang Wan, and Wenjin Wang are with the National Mobile Communications Research Laboratory, Southeast University, Nanjing 210096, China, and also with Purple Mountain Laboratories, Nanjing 211100, China (e-mail: \{wangyf, ymzhu, jywan, wangwj\}@seu.edu.cn).}
\thanks{Yuchen Zhang is with the Computer, Electrical, and Mathematical Science \& Engineering Division, King Abdullah University of Science and Technology, Thuwal 23955-6900, Saudi Arabia (e-mail: yuchen.zhang@kaust.edu.sa).}
\thanks{Symeon Chatzinotas, and Björn Ottersten are with the Interdisciplinary Centre for Security, Reliability and Trust (SnT), University of Luxembourg (E-mails: \{symeon.chatzinotas, bjorn.ottersten\}@uni.lu).}
}

\maketitle

\begin{abstract}
Symbol-level precoding (SLP) improves multiuser transmission by exploiting constructive interference based on user symbols. A microwave linear analog computer (MiLAC) maps a reduced number of radio-frequency (RF) port input signals to a large antenna array through a tunable passive microwave network. Their combination enables MiLAC to reduce the RF-chain burden of SLP, while SLP offers symbol-dependent design flexibility that can alleviate the performance loss caused by a limited number of RF chains and practical circuit constraints of MiLAC. This work investigates the joint design of SLP and MiLAC and analyzes how a limited number of RF chains and practical circuit constraints affect performance. 
For ideal MiLAC networks, we characterize the condition for attaining fully digital SLP performance and quantify the associated performance loss otherwise. We propose the joint effective subspace MiLAC-SLP (JES-MiLAC-SLP) algorithm to jointly optimize the analog beamforming matrix and SLP signals over a symbol block. For stem-connected MiLAC networks with finite tuning ranges and quantized tuning states, we propose the hardware-constrained JES-MiLAC-SLP (HC-JES-MiLAC-SLP) algorithm. It identifies circuit state changes for which SLP can better compensate the errors caused by circuit limitations, and updates the circuit state accordingly. Numerical results validate the derived findings and show that joint design reduces errors relative to separate MiLAC and SLP designs. 
Under the considered configurations, HC-JES-MiLAC-SLP substantially narrows the gap to fully digital SLP.
\end{abstract}
\vspace{-3mm}

\vspace{-2mm}
\section{Introduction}
\vspace{-1mm}
\label{sec:introduction}
\IEEEPARstart{I}{n} a multiuser multiple-input single-output {(MISO)} downlink, symbol-level precoding (SLP) jointly designs the transmit signal using the channel state and transmitted symbols \cite{alodeh2018_slp_survey,li2020_slp_tutorial}. Conventional linear precoding generally suppresses multiuser interference without considering the current symbols \cite{wang2025tensor_equivariance_mu_mimo,wang2026multisat}, whereas SLP allows the target received signals to lie within their corresponding constructive interference regions (CIRs), thereby making part of the multiuser interference favorable for symbol detection. Early studies established {constructive interference (CI)} transmission for phase-shift keying (PSK) modulations \cite{masouros2015_green_signal,alodeh2015_constructive_multiuser}, and subsequent work extended it to multilevel modulation \cite{alodeh2017slp_multilevel}, showing that {SLP can turn multiuser interference into signals that aid symbol detection, thereby reducing the required transmit power or improving demodulation performance}.

Subsequent studies extended SLP in terms of optimization criteria and implementation methods. Two early fundamental SLP problems are the power-minimization criterion, which minimizes the transmit power under CI and signal-to-interference-plus-noise ratio (SINR) constraints \cite{haqiqatnejad2018_distance_ci}, and the SINR-balancing criterion, which maximizes the SINR metric under a given power \cite{li2018_closed_form_psk,haqiqatnejad2018_distance_ci,li2021_closed_form_multilevel}. Furthermore, the study in \cite{wang2022ci_wmmse} proposed a weighted mean-square error (MSE) minimization criterion for SLP, which incorporates noise statistics into the objective function and improves SLP demodulation performance over the entire {signal-to-noise ratio (SNR)} range. In addition, the symbol error rate (SER) minimization criterion directly improves symbol detection performance. 
Related extensions have considered demodulation in coded systems and channel aging \cite{wang2024soft_demodulator_slp,wang2024robust_aging_slp}. 
Regarding complexity, the works in \cite{li2018_closed_form_psk,li2021_closed_form_multilevel,domouchtsidis2019_low_complexity_slp} considered low-complexity iterative and closed-form designs, reducing the computational or hardware burden. 
{The studies in} \cite{zhang2025unlocking_slp_tenn,zhang2026accelerate_slp_tenn} proposed a tensor-equivariant deep learning method, which further substantially improves SLP efficiency. 
Although these methods adopt different criteria, they generally employ fully digital antenna architectures.
In terms of implementation, existing studies extended SLP to hybrid precoding architectures {that combine digital and analog processing} to reduce the number of {costly and power-consuming} radio-frequency (RF) chains. Early work investigated symbol-level hybrid precoding for {millimeter-wave} multiuser systems \cite{liu2019_symbol_level_hybrid}, while subsequent schemes further considered robustness against phase errors and energy efficiency \cite{hegde2019_hybrid_phase_errors,haqiqatnejad2021_hybrid_slp}. 
The study in \cite{lee2023_hybrid_mse_slp} developed a low-complexity hybrid SLP method with prescribed MSE targets and explicit CIR constraints on the scaled noiseless received signals.

Recently, the microwave linear analog computer (MiLAC) has been proposed as an architecture for antenna-array signal processing \cite{nerini2025milac_part1}. It performs linear transformations directly in the RF domain through tunable microwave networks and connects a large-scale antenna array through a reduced number of RF input ports, thereby reducing the number of RF chains and the associated hardware and computational costs \cite{nerini2025milac_part2,nerini2026milac_opportunities}. Most prior theoretical studies of MiLAC consider lossless reciprocal microwave networks whose scattering matrices are unitary and symmetric. Under this assumption, MiLAC has been shown to achieve fully digital capacity in point-to-point multiple-input multiple-output (MIMO) systems, whereas a single MiLAC generally cannot achieve fully digital performance in multiuser systems, although the performance gap diminishes as the number of antennas grows \cite{nerini2025milac_capacity,fang2026lossless_reciprocal_milac,wu2026milac_miso_limits}.
To overcome this limitation, existing studies combine a low-dimensional digital precoder with MiLAC or employ a two-layer MiLAC structure to obtain performance comparable to fully digital beamforming under the respective dimensional and network conditions \cite{wu2026milac_miso_limits,zhou2026_two_layer_milac}.
Beyond communication performance, conditions for attaining the fully digital Cram\'{e}r--Rao bound with MiLAC have also been characterized for direction-of-arrival estimation \cite{zhang2026milac_crb}.
To reduce circuit complexity, stem-connected networks use structured connections to reduce the number of tunable components while retaining capacity-achieving performance in point-to-point MIMO systems \cite{nerini2026reduced_complexity_milac}. In multiuser systems, stem-connected MiLAC combined with digital precoding can also realize fully digital beamforming. Existing studies further consider bounded and quantized susceptance values and develop corresponding beamforming optimization methods \cite{zhang2026stem_connected_milac}.
Energy efficiency has also been investigated for MiLAC beamforming under digital-to-analog converter quantization noise, together with its tradeoff with spectral efficiency \cite{zhang2026milac_ee}.

Although MiLAC and SLP have been studied separately, the theoretical foundations of their joint design remain largely unexplored. 
{Existing communication studies primarily focus on capacity, sum-rate, and energy efficiency optimization of linear precoding}, whereas the performance limits of joint MiLAC-SLP design remain to be explored, including the conditions for attaining fully digital SLP performance, the performance losses caused by insufficient RF chains or hardware constraints such as bounded and quantized susceptance values, and the potential to reduce these losses by exploiting CI.
The potential of joint design arises from two aspects. Although the analog beamforming matrix realized by MiLAC remains fixed within a symbol block, its design can exploit both CSI and block-level multiuser symbol information. SLP allows the target signals to be adjusted within the CIR, providing additional design degrees of freedom to alleviate the limitations imposed by MiLAC.
In this paper, we systematically investigate the joint design of MiLAC and SLP, reveal the impact of MiLAC on SLP performance and the potential gains of their integration, and develop joint algorithms that coordinate MiLAC optimization with symbol-level transmission. The main contributions are summarized as follows.
{
\begin{itemize}
\item We establish a joint MiLAC-SLP design framework under an antenna-side power constraint to minimize the MSE between the received signals and the targets within the CIR by jointly optimizing the block-fixed MiLAC network, symbol-level digital signals, and target signals. The framework exploits both CSI and block-level user-symbol information in analog network design, with adjustable CIR targets mitigating the performance loss caused by limited RF chains and circuit constraints.
\item For ideal lossless reciprocal MiLAC networks, we characterize the performance limits of joint design. Under the stated conditions, including a full-row-rank channel, fully digital SLP performance is attainable when the number of RF chains is no smaller than the number of users. With fewer RF chains, we derive a residual lower bound and relate the high-SNR limit to the eigenvalue tail sum of the covariance of feasible target signal blocks. 
We further reformulate network design equivalently as low-dimensional effective-channel subspace optimization and propose JES-MiLAC-SLP, which alternates between {optimal subspace updates for fixed CIR targets} and {joint optimization of the digital signals and CIR targets for a fixed subspace}.
\item For stem-connected MiLAC networks with limited susceptance ranges and finite quantization resolution, we formulate the hardware-constrained joint optimization problem and derive a performance perturbation bound to support circuit initialization from the ideal solution. We then propose hardware-constrained JES-MiLAC-SLP (HC-JES-MiLAC-SLP), which accounts for local adjustments of the digital signals and CIR targets when determining the circuit update direction, guiding the selection of improved candidate updates to promote rapid reduction of the joint objective.
\item Numerical results validate the theoretical analysis and show that JES-MiLAC-SLP substantially outperforms channel-based MiLAC design, approaching fully digital SLP performance even with fewer RF chains than users. For hardware-constrained networks, HC-JES-MiLAC-SLP further reduces MSE and SER compared with directly mapping the ideal design to hardware by jointly optimizing the quantized circuit state and SLP signals.
\end{itemize}
}

Section~\ref{sec:system} presents the system model. Sections~\ref{sec:lossless} and~\ref{sec:physical} develop the general lossless reciprocal and hardware-constrained stem-connected designs, respectively. Section~\ref{sec:numerical} presents numerical results, and Section~\ref{sec:conclusion} concludes the paper.

{\textit{Notation}: Bold lowercase/uppercase letters denote vectors/matrices; superscripts $T,H,\dagger$ denote transpose, conjugate transpose, and Moore-Penrose pseudoinverse. $\mat I_n$ is the $n$th-order identity; $\operatorname{tr}$, $\rank$, and $\Range$ denote trace, rank, and column space. $\|\cdot\|_F$ is the Frobenius norm; $\|\cdot\|_2$ is the Euclidean/spectral norm for vectors/matrices. For Hermitian matrices, $\mat X\preceq\mat Y$ (i.e., $\mat Y\succeq\mat X$) means $\mat Y-\mat X$ is positive semidefinite; $\mat X\succ\mat0$ means positive definite. Vector inequalities are elementwise. For $\mat X\succ\mat0$, $\mat X^{1/2}$ is its unique Hermitian positive definite square root and $\mat X^{-1/2}=(\mat X^{1/2})^{-1}$. $\operatorname{diag}(a_1,\ldots,a_n)$ forms a diagonal matrix with the listed entries; $\Re\{\cdot\}$ and $\Im\{\cdot\}$ denote real and imaginary parts, and $j^2=-1$. $\mathbb E$ denotes expectation, with subscripts identifying random quantities; $|\mathcal A|$ denotes cardinality and $[\mat X]_{i,j}$ a matrix entry. $\to$, $\downarrow$, and $\mapsto$ denote convergence, convergence from above, and mapping.}

\section{MiLAC Architecture and Constructive-Interference Transmission}
\label{sec:system}
\subsection{MiLAC Architecture}
\label{subsec:architecture_signal_model}

\begin{figure*}[!t]
\centering
\includegraphics[width=0.98\textwidth]{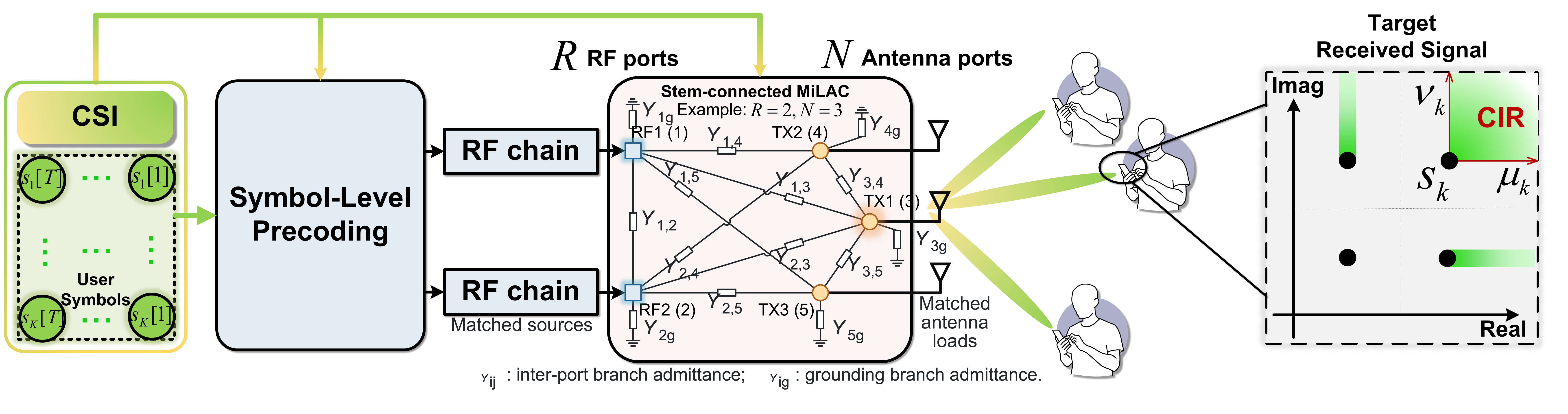}
\vspace{-3mm}
\caption{MiLAC-aided multiuser downlink with SLP. {The circuit illustrates a stem-connected network with two RF ports and three antenna ports. The right inset shows quadrature amplitude modulation (QAM) constellation points and their CIRs. The system model applies to $1\leq R\leq N$, without restricting the relation between $R$ and $K$. }}
\label{fig:system_model}
\vspace{-4mm}
\end{figure*}

As illustrated in \figref{fig:system_model}, an $N$-antenna base station serves $K$ single-antenna users through $R$ RF input ports, where $1\leq R\leq N$. SLP designs the digital signals using the user symbols and CSI. {Each transmit branch converts the digital SLP output into an RF-port input signal through digital-to-analog conversion, filtering, upconversion, and power amplification before the MiLAC \cite{nerini2026milac_opportunities}.} Its tunable microwave network combines the RF-port input signals into $N$ antenna transmit signals. {The channel $\mat H\in\mathbb C^{K\times N}$ from the BS antenna ports to the users is assumed to remain constant over $T$ symbols and to be perfectly known at the transmitter.}
{Channel estimation can be performed by switching MiLAC configurations during training and jointly processing multiple observations \cite{zhang2026milac_ce}, a procedure beyond the scope of this work.}
We first introduce the modeling of MiLAC networks.
\subsubsection{Lossless Reciprocal Model}
\label{subsubsec:general_milac_model}
{As joint MiLAC-SLP design is still at an early stage of research, we adopt a lossless network model composed of ideal reactive components, consistent with existing MiLAC studies \cite{nerini2025milac_part1,fang2026lossless_reciprocal_milac,wu2026milac_miso_limits}. This model enables us to analyze how the number of RF chains, network topology, and susceptance tuning constraints affect transmission performance, providing a theoretical basis for joint algorithm design and subsequent hardware implementation.}
Let $Y_{p,q}$ and $Y_{pg}$ denote the admittances of the branch between ports $p,q$ and the branch from port $p$ to ground, respectively, as labeled in \figref{fig:system_model}. These branches determine the network admittance matrix $\mat Y_{\rm net}\in\mathbb C^{(N+R)\times(N+R)}$, which maps port voltages to currents entering the network. Losslessness means that the network dissipates no average power and implies $\mat Y_{\rm net}=j\mat B_{\rm net}$ for a Hermitian susceptance matrix $\mat B_{\rm net}$. Reciprocity requires $\mat Y_{\rm net}^T=\mat Y_{\rm net}$, under which $\mat B_{\rm net}$ is real and symmetric. For this matrix, the off-diagonal entries are the negative inter-port branch susceptances, while each diagonal entry aggregates the incident inter-port and shunt-to-ground susceptances \cite{nerini2025milac_part1,nerini2026reduced_complexity_milac,zhang2026stem_connected_milac}.
The scattering matrix $\boldsymbol\Theta$ maps incident signals to outgoing signals. For real reference impedance $Z_0>0$ and reference admittance $Y_0=1/Z_0$, it is related to the circuit matrices by \cite{nerini2025milac_part1,fang2026lossless_reciprocal_milac}
\begin{equation}
\begin{aligned}
\boldsymbol\Theta
=(Y_0\mat I-j\mat B_{\rm net})(Y_0\mat I+j\mat B_{\rm net})^{-1}.
\end{aligned}
\label{eq:cayley_network}
\end{equation}
The resulting scattering matrix is unitary and symmetric \cite{nerini2025milac_part1}:
\begin{equation}
\boldsymbol\Theta^H\boldsymbol\Theta=\mat I_{N+R},
\qquad
\boldsymbol\Theta^T=\boldsymbol\Theta.
\label{eq:lossless_reciprocal}
\end{equation}
Ordering the RF ports before the antenna ports yields the following partitioned scattering relation. The diagonal blocks describe reflection and coupling within each port group, whereas the off-diagonal blocks describe transmission between the two groups.
Let $\vect w_{\rm RF}^{+}$ and $\vect w_{\rm ant}^{+}$ denote the power waves incident on the RF and antenna ports, respectively, and let $\vect w_{\rm RF}^{-}$ and $\vect w_{\rm ant}^{-}$ denote the corresponding outgoing waves \cite{kurokawa1965power_waves}. Their scattering relation is
\begin{equation}
\begin{bmatrix}
\vect w_{\rm RF}^{-}\\
\vect w_{\rm ant}^{-}
\end{bmatrix}
=
\boldsymbol\Theta\begin{bmatrix}
\vect w_{\rm RF}^{+}\\
\vect w_{\rm ant}^{+}
\end{bmatrix}
=
\begin{bmatrix}
\boldsymbol\Theta_{11} & \boldsymbol\Theta_{12}\\
\boldsymbol\Theta_{21} & \boldsymbol\Theta_{22}
\end{bmatrix}
\begin{bmatrix}
\vect w_{\rm RF}^{+}\\
\vect w_{\rm ant}^{+}
\end{bmatrix}.
\label{eq:partitioned_scattering_relation}
\end{equation}
When the antenna loads are matched to $Z_0$, $\vect w_{\rm ant}^{+}=\vect0$, and hence $\vect w_{\rm ant}^{-}=\boldsymbol\Theta_{21}\vect w_{\rm RF}^{+}$.
{With matched antenna loads, unitarity ensures that the RF incident power is split between antenna-port output and reflection back to the RF ports. Therefore, the total antenna-port output power cannot exceed the total RF-port incident power.}

\subsubsection{Connection Topology}
\label{subsubsec:connection_topology}
\label{subsubsec:stem_topology}
The connection topology specifies the available tunable branches and hence the structure of $\mat B_{\rm net}$.
Let $\vect b=[b_1,\ldots,b_{N_{\rm c}}]^T\in\mathbb R^{N_{\rm c}}$ collect the physical branch susceptances, namely the imaginary parts of their admittances, and let $\vect e_p$ be the standard basis vector for port $p$. For branch $i$, the connection vector is $\vect v_i=\vect e_p-\vect e_\ell$ if it joins ports $p,\ell$, and $\vect v_i=\vect e_p$ if it joins port $p$ to ground.
Summing the contributions of all branches gives \cite{nerini2025milac_part1}
\begin{equation}
\textstyle
\mat B_{\rm net}(\vect b)=\sum_{i=1}^{N_{\rm c}}b_i\vect v_i\vect v_i^T.
\label{eq:network_susceptance}
\end{equation}
The stem-connected topology is a typical low-complexity topology that reduces the number of tunable components \cite{nerini2026reduced_complexity_milac,zhang2026stem_connected_milac}. It selects the central-port set $\mathcal C=\{1,\ldots,2R-1\}$, which includes all RF ports. The central ports are mutually connected, each remaining port connects only to the central ports, and every port has a grounding branch. Thus, $[\mat B_{\rm net}]_{p\ell}=0$ for distinct noncentral ports $p,\ell\notin\mathcal C$. In the $R=2$, $N=3$ example in \figref{fig:system_model}, RF ports 1 and 2 and antenna port 3 are central, whereas ports 4 and 5 have no direct connection.
This topology has $N_{\rm c}=R(2N+1)$ adjustable susceptances \cite{nerini2026reduced_complexity_milac,zhang2026stem_connected_milac}. For any real susceptance vector, \eqref{eq:network_susceptance} constructs a real symmetric matrix, so the resulting $\boldsymbol{\Theta}$ satisfies \eqref{eq:lossless_reciprocal}.

\subsubsection{Finite-Range and Quantized Susceptances}
\label{subsubsec:discrete_susceptances}
For a given connection topology, each susceptance in ${\bf b}$ can be restricted to the following finite-range, $q$-bit quantized values \cite{zhang2026stem_connected_milac}:
\begin{align}
&b_i\in\mathcal Q_q
=\{-B_{\max},-B_{\max}+\Delta_q,\ldots,B_{\max}\},
\label{eq:quantized_susceptance}\\
&\Delta_q=2B_{\max}/(2^q-1),
\label{eq:quantization_spacing}
\end{align}
where $B_{\max}>0$ specifies the susceptance range, and $\Delta_q$ is the spacing between adjacent available values. 
Because the quantized susceptances remain real, \eqref{eq:network_susceptance} remains real symmetric and preserves the lossless and
reciprocal conditions \eqref{eq:lossless_reciprocal}.

These models provide the basis for the theoretical analysis and joint design of MiLAC and SLP under different network constraints.

\subsection{MiLAC-SLP Transmission Model}
{To ease tuning and control requirements, we fix the MiLAC network within each block and design RF-port input signals per symbol.} Following \eqref{eq:partitioned_scattering_relation}, define
\begin{equation}
\FM\triangleq\boldsymbol\Theta_{21}\in\mathbb C^{N\times R},
\ \ 
\vect u[t]\triangleq\vect w_{\rm RF}^{+}[t],
\ \ 
\vect x[t]\triangleq\vect w_{\rm ant}^{-}[t],
\label{eq:milac_signal_mapping}
\end{equation}
where $\vect u[t]\in\mathbb C^R$ is the RF-port input signal supplied by the RF chains, and $\vect x[t]\in\mathbb C^N$ is the output signal delivered to the antenna ports. Their squared norms represent the incident and antenna-port output powers, respectively. Assuming that the source impedances and antenna loads are matched to the reference impedance $Z_0>0$, the antenna-port output satisfies $\vect x[t]=\FM\vect u[t]$, and the received signal is
\footnote{{Let $\vect v_{\rm s}$ denote the peak complex open-circuit voltages of the RF sources with series impedance $Z_0$, and let $\vect v_{\rm out}$ denote the voltages across the matched antenna loads. Since the source and port reference impedances are both $Z_0$, the sources launch the incident voltage $\vect v_{\rm s}/2$ \cite{kurokawa1965power_waves}. Hence, matched antenna loads give $\vect v_{\rm out}=\boldsymbol\Theta_{21}\vect v_{\rm s}/2$ \cite{nerini2025milac_part1,fang2026lossless_reciprocal_milac}. We scale the port signals as $\vect u=\vect v_{\rm s}/(2\sqrt{2Z_0})$ and $\vect x=\vect v_{\rm out}/\sqrt{2Z_0}$, yielding $\vect x=\boldsymbol\Theta_{21}\vect u$. Their squared norms equal the RF incident power and antenna-port output power, respectively. Therefore, we adopt the same antenna-side power budget as conventional linear precoding and SLP for comparison.}}
\begin{equation}
\vect x[t]=\FM\vect u[t],
\qquad
\vect y[t]=\mat H\vect x[t]+\vect n[t],
\label{eq:signal_model}
\end{equation}
{where $\vect n[t]\sim\CN(\vect0,\sigma^2\mat I_K)$ is additive white Gaussian noise.}
{The analog beamforming matrix is fixed throughout the $T$-symbol block, whereas $\vect u[t]$ is optimized for the user symbols at each time. Stacking the signals as columns, define the received signal block $\mat Y=[\vect y[1],\ldots,\vect y[T]]\in\mathbb C^{K\times T}$, the RF-port input-signal block $\mat U=[\vect u[1],\ldots,\vect u[T]]\in\mathbb C^{R\times T}$, and the noise block $\mat N=[\vect n[1],\ldots,\vect n[T]]\in\mathbb C^{K\times T}$. The corresponding noiseless received signal block $\widetilde{\mat Y}\in\mathbb C^{K\times T}$ satisfies}
{
\begin{equation}
\mat Y=\widetilde{\mat Y}+\mat N,
\qquad
\widetilde{\mat Y}=\mat H\FM\mat U.
\label{eq:received_signal_blocks}
\end{equation}
}
For $\vect u[t]=\mat W\vect s[t]$ with a blockwise fixed $\mat W$, this architecture reduces to the established hybrid digital-MiLAC linear-beamforming architecture \cite{wu2026milac_miso_limits}. {The average power delivered to the antenna ports is constrained as}
\begin{equation}
\left\|\FM\mat U\right\|_F^2/T\leq P_{\rm T}.
\label{eq:antenna_power}
\end{equation}
{We adopt an average antenna-port power constraint to provide a common transmit power budget for different architectures. In practical implementations, suitable power-amplifier operating points can be selected based on the signal peaks in each RF branch, or signal peaks and power differences among branches can be further controlled to reduce nonlinear amplitude and phase distortion \cite{spano2018nonlinear_slp}.}
Existing MiLAC studies also consider the source signal power constraint \cite{wu2026milac_miso_limits}, Remark~\ref{rem:source_power} explains how the theory and algorithm in Section~\ref{sec:lossless} apply under this constraint.

\vspace{-2mm}
\subsection{CIR Characterization}
\label{subsec:ci_targets}

The right inset of \figref{fig:system_model} illustrates a {16-QAM} example, where black dots denote the original constellation points and green regions show the CIRs of the outer constellation points. SLP designs the transmit signal using the known user symbols and allows the receive target to be selected within the corresponding CIR to exploit CI \cite{alodeh2017slp_multilevel,wang2022ci_wmmse}. At symbol time $t$, let $s_k[t]$ and $\widetilde{s}_k[t]$ denote the intended symbol and receive target of user $k$, respectively, and let $\mathcal D_k[t]$ denote the CIR associated with $s_k[t]$. The target satisfies
\begin{equation}
\widetilde{s}_k[t]\in\mathcal D_k[t],
\qquad k=1,\ldots,K,\quad t=1,\ldots,T,
\label{eq:ci_target_constraint}
\end{equation}
{where $\widetilde{s}_k[t]$ is a design target, whereas $[\widetilde{\mat Y}]_{k,t}$ in \eqref{eq:received_signal_blocks} is the actual noiseless received signal. The subsequent MSE minimization makes the receive-scaled signal approximate the selected target. Collecting these targets in $\widetilde{\mat S}$ gives the block constraint $\widetilde{\mat S}\in\mathcal D(\mat S)$.}
Let $\mu_k[t]$ and $\nu_k[t]$ denote the two CIR boundary directions associated with the intended symbol $s_k[t]$. The two red arrows starting from $s_k$ in \figref{fig:system_model} illustrate these directions for a corner symbol; the time index $t$ is omitted in the figure. Using these directions, the CIR is parameterized as \cite{wang2022ci_wmmse}
\begin{equation}
\begin{aligned}
\mathcal D_k[t]=\bigl\{\widetilde s_k[t]:
&\ \widetilde s_k[t]=s_k[t]
+\delta_{\mu,k}[t]\mu_k[t]\\
&+\delta_{\nu,k}[t]\nu_k[t],\quad
\delta_{\mu,k}[t],\delta_{\nu,k}[t]\geq0\bigr\}.
\end{aligned}
\label{eq:cir_boundary_description}
\end{equation}
{The intended symbol determines the boundary directions, while the nonnegative coefficients determine the target's position within the CIR.}
{For phase-shift keying (PSK) symbols, the CIR extends from each constellation point along directions parallel to its two decision boundaries.} In contrast, for QAM symbols, the CIR of an interior point reduces to the point itself, whereas only the constellation points on the outermost ring have extendable CIRs \cite{alodeh2017slp_multilevel,wang2022ci_wmmse}.
{Let $\boldsymbol\Lambda[t]\in\mathbb R^{2K\times2K}$ be the CIR boundary matrix constructed from $\mu_k[t]$ and $\nu_k[t]$ as in \cite{wang2022ci_wmmse}, and define $\boldsymbol\delta[t]=[\boldsymbol\delta_\mu[t]^T,\boldsymbol\delta_\nu[t]^T]^T$. Then, \eqref{eq:ci_target_constraint} is equivalently expressed as}
\begin{equation}
\widehat{\vect{s}}[t]
=
\overline{\vect{s}}[t]+
\boldsymbol\Lambda[t]\boldsymbol\delta[t],
\qquad
\boldsymbol\delta[t]\succeq\vect0.
\label{eq:ci_region}
\end{equation}
where $\vect{s}[t]=[s_1[t],\ldots,s_K[t]]^T$, with the real representations  $\overline{\vect{s}}[t]=[\Re\{\vect{s}[t]\}^T,\Im\{\vect{s}[t]\}^T]^T$ and $\widehat{\vect{s}}[t]=[\Re\{\widetilde{\vect{s}}[t]\}^T,\Im\{\widetilde{\vect{s}}[t]\}^T]^T$.
 

\section{Ideal MiLAC-SLP Design}
\label{sec:lossless}
Reducing the number of RF chains helps lower hardware costs but may affect multiuser transmission performance. This section therefore investigates how the number of RF chains affects the performance of lossless reciprocal MiLAC-SLP. We first formulate the joint optimization problem, analyze the conditions for attaining fully digital performance, and characterize the performance limitations when RF chains are insufficient. Based on these results, we equivalently reformulate the MiLAC network optimization as the optimization of a low-dimensional effective channel basis matrix, and design the initialization and optimization steps of JES-MiLAC-SLP.

\subsection{Joint MiLAC-SLP Optimization Problem}
\label{subsec:lr_problem}
We first consider a lossless reciprocal MiLAC network satisfying \eqref{eq:lossless_reciprocal}, without additional restrictions on its connection topology or susceptances.
In particular, the upper-left block obeys
\begin{equation}
\boldsymbol\Theta_{11}^H\boldsymbol\Theta_{11}+\FM^H\FM=\mat I_R.
\label{eq:forward_contraction}
\end{equation}
Thus, $\FM^H\FM\preceq\mat I_R$ follows directly from such model.
{We consider an MSE minimization problem that jointly optimizes the analog beamforming matrix $\FM$, RF-domain signal block $\mat U$, and target signals $\widetilde{\mat S}$ within the CIR to minimize the MSE between the received signals and these targets, thereby improving demodulation performance \cite{wang2022ci_wmmse, wang2024robust_aging_slp, zhang2025unlocking_slp_tenn}.}
\begin{subequations}
\label{eq:lr_joint_problem}
\begin{align}
\mathcal P_1:\quad
&\min_{\FM,\mat U,\widetilde{\mat S},\gamma>0}\quad
\frac{1}{KT}\mathbb E_{\mat N}\left\|
\frac{\mat H\FM\mat U+\mat N}{\gamma}-\widetilde{\mat S}
\right\|_F^2 \label{eq:lr_joint_objective}\\
&\quad\text{s.t.}\quad
\FM^H\FM\preceq\mat I_R,\quad \eqref{eq:antenna_power},\quad
\widetilde{\mat S}\in\mathcal D(\mat S),
\end{align}
\end{subequations}
{where $\gamma>0$ is a receive scaling factor optimized at the BS and obtained at the receivers through BS signaling or estimation to adjust the amplitude scale of the received signals before demodulation \cite{wang2022ci_wmmse,wang2024soft_demodulator_slp}.}
{The coupling between the MiLAC network and the SLP variables makes this joint optimization problem nonconvex.}


We next show that the contraction $\FM^H\FM\preceq\mat I_R$ in the lossless reciprocal model can be replaced, without changing the optimum, by a column-orthonormal analog beamforming matrix.

\begin{lemma}[Equivalent Representation of the Analog Beamforming Matrix]
\label{lem:representation}
{For $1\leq R\leq N$,} under the power constraint in \eqref{eq:antenna_power}, the optimum of \eqref{eq:lr_joint_problem} is unchanged if the contraction constraint $\FM^H\FM\preceq\mat I_R$ is replaced by
\begin{equation}
\FM^H\FM=\mat I_R.
\label{eq:orthonormal_representative}
\end{equation}
In particular, for every feasible $\FM$ and $\mat U$, there exist $\bar{\mat F}_{\mathrm M}$ and $\bar{\mat U}$ satisfying \eqref{eq:orthonormal_representative} such that $\bar{\mat F}_{\mathrm M}\bar{\mat U}=\FM\mat U$.
\end{lemma}
\begin{proof}
See Appendix~\ref{app:proof_representation}.
\end{proof}
\lemref{lem:representation} permits the subsequent analysis to use a column-orthonormal analog beamforming matrix without changing the optimum of $\mathcal P_1$. When the relevant construction conditions are satisfied \cite{zhang2026stem_connected_milac}, the matrix $\FM$ can be realized by a stem-connected network in Section~\ref{subsubsec:connection_topology}, preserving the transmission performance of the design in this section \cite{nerini2026reduced_complexity_milac,zhang2026stem_connected_milac}.

\subsection{SLP Performance Limits With Fewer RF Chains}
\label{subsec:lr_theory}

{Using fully digital SLP as a benchmark, we analyze how the number of RF chains affects the achievable performance, identifying when fully digital performance can be preserved and what limitations arise otherwise. This analysis motivates the subsequent joint design of MiLAC and SLP in the RF-chain-limited regime, where $R\leq K$.}

Let $f_{\rm FD}^\star$ denote the minimum objective value of the fully digital version of \eqref{eq:lr_joint_problem}, obtained by allowing an optimal antenna-domain transmit block $\mat X\in\mathbb C^{N\times T}$ under \eqref{eq:antenna_power}. 
For a feasible analog beamforming matrix $\FM$, let $\mat P_{\mat H\FM}$ be the orthogonal projector onto $\Range(\mat H\FM)$, so that $\mat P_{\mat H\FM}\vect y$ retains the part of any $\vect y\in\mathbb C^K$ lying in this column space, while $(\mat I_K-\mat P_{\mat H\FM})\vect y$ is perpendicular to it. Let $f(\FM)$ denote the minimum value of \eqref{eq:lr_joint_problem} over $\mat U$, $\widetilde{\mat S}$, and $\gamma$ for fixed $\FM$.

\begin{lemma}[RF-Chain Threshold and Residual Bound]
\label{prop:loss_regime}
Assume $\rank(\mat H)=K$ and $R\leq N$.

(i) If $R\geq K$, there exists a feasible lossless reciprocal MiLAC network such that $f(\FM)=f_{\rm FD}^\star$.

(ii) If $R<K$, then $\rank(\mat H\FM)<K$ for every feasible $\FM$, and
\begin{equation}
f(\FM)\geq
\frac{1}{KT}
\min_{\widetilde{\mat S}\in\mathcal D(\mat S)}
\left\|(\mat I_K-\mat P_{\mat H\FM})\widetilde{\mat S}\right\|_F^2.
\label{eq:rank_deficient_bound}
\end{equation}
\end{lemma}

\begin{proof}
See Appendix~\ref{app:proof_loss_regime}.
\end{proof}
\begin{remark}
{For $\mathcal P_1$, \lemref{prop:loss_regime} distinguishes two cases. When $R\geq K$, constructing a column-orthonormal analog beamforming matrix from the channel's right singular vectors to contain $\Range(\mat H^H)$ and solving conventional SLP on $\mat H\FM$ attain the fully digital optimum \cite{wu2026milac_miso_limits,wang2022ci_wmmse}.}
Operating with $R<K$ allows fewer RF chains and lower hardware costs for a given number of users, or simultaneous transmission to more users with a fixed RF-chain configuration \cite{garcia2019_more_users,li2021_closed_form_multilevel}. The effective channel {$\mat H\FM$} then has rank at most $R$, limiting the available spatial degrees of freedom and potentially causing the performance loss characterized by \eqref{eq:rank_deficient_bound}. Therefore, the subsequent study focuses on characterizing the performance limits and jointly designing MiLAC and SLP to improve transmission performance for $R\leq K$.
\end{remark}

\vspace{-2mm}
\subsection{Reformulation of the CI Optimization}
\label{subsec:ci_reformulation}

With the preceding column-orthonormal representation of the analog beamforming matrix, we further simplify the optimization of the SLP-related variables {$\mat U$, $\widetilde{\mat S}$, and $\gamma$} in $\mathcal P_1$.
{Let $\mat V=\mat U/\gamma$, so that the rescaled noiseless received signal is $\mat H\FM\mat V$. For fixed $\FM$, $\mat V$, and $\widetilde{\mat S}$, the optimal solution uses the full transmit power budget, i.e., $\|\FM\mat U\|_F^2/T=P_{\rm T}$. Therefore,}
\begin{equation}
\gamma^\star=\sqrt{{P_{\rm T}T}/{\|\mat V\|_F^2}},\qquad \mat U^\star=\gamma^\star\mat V.
\label{eq:ci_scaling_recovery}
\end{equation}
Substituting \eqref{eq:ci_scaling_recovery} into \eqref{eq:lr_joint_problem} eliminates $\mat U$ and $\gamma$, and yields the following problem, where $\xi\triangleq K\sigma^2/P_{\rm T}$.
\begin{subequations}
\label{eq:orthonormal_joint_problem}
\begin{align}
\mathcal P_2:\quad
\min_{\FM,\,\mat V,\,\widetilde{\mat S}}\quad
&\frac{1}{KT}\left(\|\mat H\FM\mat V-\widetilde{\mat S}\|_F^2+\xi\|\mat V\|_F^2\right)\\
\text{s.t.}\quad
&\FM^H\FM=\mat I_R,\\
&\widetilde{\mat S}\in\mathcal D(\mat S).
\end{align}
\end{subequations}
{For fixed $\FM$ and $\widetilde{\mat S}$, $\mathcal P_2$ in \eqref{eq:orthonormal_joint_problem} is an unconstrained quadratic optimization in $\mat V$, whose optimal solution is}
\begin{equation}
\mat V^\star=\left[(\mat H\FM)^H\mat H\FM+\xi\mat I_R\right]^{-1}(\mat H\FM)^H\widetilde{\mat S}.
\label{eq:normalized_rf_update}
\end{equation}
{Substituting $\mat V^\star$ from \eqref{eq:normalized_rf_update} into the objective gives a minimum value expressed using the following weight matrix.}
\begin{equation}
\mat W_{\mat H\FM}
=\mat I_K-\mat H\FM
\left[(\mat H\FM)^H\mat H\FM+\xi\mat I_R\right]^{-1}
(\mat H\FM)^H.
\label{eq:reduced_mse_weight}
\end{equation}
{The original joint problem is thus equivalently reduced to the following optimization over only $\FM$ and $\widetilde{\mat S}$.}
\begin{align}
\mathcal P_3:\quad
\min_{\substack{\FM,\,\widetilde{\mat S}\in\mathcal D(\mat S)}}\quad
&\frac{1}{KT}\operatorname{tr}\!\left(
\widetilde{\mat S}^H\mat W_{\mat H\FM}\widetilde{\mat S}
\right)\nonumber\\
\text{s.t.}\quad
&\FM^H\FM=\mat I_R.
\label{eq:reduced_f_problem}
\end{align}

\subsection{Effective Channel Subspace Transformation}
\label{subsec:subspace_formulation}
In the objective functions of $\mathcal P_2$ in \eqref{eq:orthonormal_joint_problem} and $\mathcal P_3$ in \eqref{eq:reduced_f_problem}, $\FM$ affects the objective only through $\mat H\FM$. {We introduce the effective channel basis matrix $\mat A\in\mathbb C^{K\times R}$ with $\mat A^H\mat A=\mat I_R$ to represent the $R$-dimensional effective channel subspace to be optimized.}
For fixed $\FM$, we define $f(\FM)$ as the minimum objective value of $\mathcal P_3$ in \eqref{eq:reduced_f_problem} over $\widetilde{\mat S}$.
{The following proposition gives a construction of the analog beamforming matrix from $\mat A$ and proves that replacing the optimization variable $\FM$ with $\mat A$ preserves the optimal value of the problem.}

\begin{proposition}[Low-Dimensional MiLAC Representation]
\label{prop:metric_realization}
Assume $\operatorname{rank}(\mat H)=K$ and $R\leq K$. For every feasible $\FM$ satisfying $\FM^H\FM=\mat I_R$, there exists a column-orthonormal effective channel basis matrix $\mat A\in\mathbb C^{K\times R}$ satisfying
\begin{equation}
\Range(\mat H\FM)\subseteq\Range(\mat A)
\label{eq:effective_range_inclusion}
\end{equation}
such that $f(\FM(\mat A))\leq f(\FM)$, where
\begin{equation}
\FM(\mat A)=\mat H^\dagger\mat A
\mat D_{\mat A}^{-1/2},
\label{eq:metric_realization}
\end{equation}
where $\mat D_{\mat A}\triangleq\mat A^H(\mat H\mat H^H)^{-1}\mat A$.
Consequently,
\begin{equation}
\min_{\FM^H\FM=\mat I_R}f(\FM)
=\min_{\mat A^H\mat A=\mat I_R}f(\FM(\mat A)).
\label{eq:fm_a_value_equivalence}
\end{equation}
\end{proposition}

\begin{proof}
See Appendix~\ref{app:proof_metric_realization}.
\end{proof}

\propref{prop:metric_realization} shows that the optimization over $\FM\in\mathbb C^{N\times R}$ can be replaced without loss of optimality by an optimization over the effective channel basis matrix $\mat A\in\mathbb C^{K\times R}$. The latter represents the $R$-dimensional effective channel subspace selected by MiLAC. Substituting $\mat H\FM(\mat A)=\mat A\mat D_{\mat A}^{-1/2}$ into $\mat W_{\mat H\FM(\mat A)}$ gives $\mat W_{\mat H\FM(\mat A)}=\mat I_K-\mat A(\mat I_{R}+\xi\mat D_{\mat A})^{-1}\mat A^H$. Accordingly, with $\mat P_{\mat A}=(\mat I_{R}+\xi\mat D_{\mat A})^{-1}$, the joint problem is equivalently written as {a nonconvex optimization} over $\mat A$
\begin{align}
&\mathcal P_4:\quad
\min_{\mat A^H\mat A=\mat I_{R}}
\ \widetilde f(\mat A),\nonumber\\
\widetilde f(\mat A)
&=\frac{1}{KT}
\min_{\widetilde{\mat S}\in\mathcal D(\mat S)}
\operatorname{tr}\!\left[
\widetilde{\mat S}^H(\mat I_K-\mat A\mat P_{\mat A}\mat A^H)
\widetilde{\mat S}\right].
\label{eq:subspace_value}
\end{align}
This transformation reduces the optimization variable from $\FM\in\mathbb C^{N\times R}$ to $\mat A\in\mathbb C^{K\times R}$.

{We then derive an asymptotically tight lower bound.}
\begin{proposition}[Loss Characterization]
\label{prop:tail}
Let $\widetilde f_\xi^\star$ denote the global optimum of \eqref{eq:subspace_value}. Then
\begin{equation}
\lim_{\xi\rightarrow0}\widetilde f_\xi^\star=
\frac{1}{K}
\inf_{\widetilde{\mat S}\in\mathcal D(\mat S)}
\sum_{i=R+1}^{K}
\lambda_i\!\left(
{\widetilde{\mat S}\widetilde{\mat S}^H}/{T}
\right),
\label{eq:tail_formula}
\end{equation}
where the eigenvalues are ordered nonincreasingly.
{The right-hand side of \eqref{eq:tail_formula} also provides a lower bound on optimal $\widetilde f_\xi^\star$ for any $\xi>0$.}
\end{proposition}

\begin{proof}
See Appendix~\ref{app:proof_tail}.
\end{proof}

The high-SNR limit isolates the performance loss that may remain due to limited RF chains as the noise contribution vanishes.
This limit is determined by the smallest $K-R$ eigenvalues of the covariance of the target signal block within the CIR. For $R=K$, the sum in \eqref{eq:tail_formula} is empty.

{
For a fixed CIR target block $\widetilde{\mat S}_i$, consider the following subproblem of $\mathcal P_4$:
\begin{align}
\min_{\mat A^H\mat A=\mat I_R}\frac{1}{KT}\operatorname{tr}\!\left[
\widetilde{\mat S}_i^H
(\mat I_K-\mat A\mat P_{\mat A}\mat A^H)
\widetilde{\mat S}_i\right].
\label{eq:fixed_target_problem}
\end{align}
Let $\mat H=\mat L_H\boldsymbol\Sigma_H\mat J_H^H$ be the thin SVD of the channel, with $\boldsymbol\Sigma_H=\operatorname{diag}(d_1,\ldots,d_K)$, and define
\begin{align}
\mat B_\xi&=\operatorname{diag}(
d_k/\sqrt{d_k^2+\xi})_{k=1}^{K},\ 
\mat M_i=\mat B_\xi\mat L_H^H\widetilde{\mat S}_i.
\label{eq:fixed_target_weighted_matrix}
\end{align}
Write the full SVD of $\mat M_i$ with singular values ordered nonincreasingly, and define
\begin{equation}
\mat M_i=\mat E_i\boldsymbol\Sigma_i\mat Z_i^H,
\qquad
\mat E_{i,R}=[\mat E_i]_{:,1:R},
\label{eq:target_subspace_basis}
\end{equation}
where $\mat E_i\in\mathbb C^{K\times K}$ is unitary{, and $[\mat E_i]_{:,1:R}$ denotes its first $R$ columns}.

\begin{proposition}[Optimal Effective Channel Subspace]
\label{prop:exact_target}
Assume $\rank(\mat H)=K$. A globally optimal solution to \eqref{eq:fixed_target_problem} is
\begin{equation}
\mat A^{\star}=\mat L_H\mat B_\xi\mat E_{i,R}
\left(\mat E_{i,R}^H\mat B_\xi^2\mat E_{i,R}\right)^{-1/2}.
\label{eq:exact_subspace_update}
\end{equation}
\end{proposition}

\begin{proof}
See Appendix~\ref{app:proof_exact_target}.
\end{proof}
}

\subsection{JES-MiLAC-SLP Algorithm}
\label{subsec:jes_algorithm}

{For $R\leq K$, the preceding equivalent reformulation transforms the optimization of $\FM$ into the optimization of the effective channel basis matrix $\mat A$. We therefore propose the JES-MiLAC-SLP algorithm, which uses an initialization motivated by the target signal covariance analysis in \propref{prop:tail} and {{alternates a globally optimal update of $\mat A$ for fixed CIR targets $\widetilde{\mat S}$ with optimization of $\widetilde{\mat S}$ and recovery of the optimal $\mat V$ for fixed $\mat A$}}. When $R=K$, $\mat A$ spans the entire space $\mathbb C^K$, and the algorithm reduces to SLP optimization with this space fixed.}

\subsubsection{Symbol Covariance-Based Initialization}
\label{subsubsec:jes_initialization}


The initialization is guided by \propref{prop:tail}. We use the original symbol block $\mat S$ to form $\mat C_0=\mat S\mat S^H/T$. Since \propref{prop:tail} relates the high-SNR loss to the energy of the target block outside the column space of $\mat A$, the initialization selects $\mat A$ to retain as much energy of $\mat S$ as possible:
\begin{equation}
\min_{\mat A^H\mat A=\mat I_R}
\frac{1}{KT}\left\|(\mat I_K-\mat A\mat A^H)\mat S\right\|_F^2,
\label{eq:jes_initialization}
\end{equation}
By the Ky Fan maximum principle \cite{horn2012matrix_analysis}, an optimal solution to \eqref{eq:jes_initialization} is $\mat A_0=[\vect e_{0,1},\ldots,\vect e_{0,R}],$
where $\vect e_{0,r}$ is a unit eigenvector of $\mat C_0$ corresponding to its $r$-th largest eigenvalue, and the selected eigenvectors are mutually orthogonal.

\subsubsection{Constructive Interference-Based Digital Signal Update}

{
For fixed $\mat A_i$, let $\mat G_H=(\mat H\mat H^H)^{-1}$ and define
\begin{align}
\mat D_i&=\mat A_i^H\mat G_H\mat A_i, &
\mat P_i&=(\mat I_R+\xi\mat D_i)^{-1},\nonumber\\
\mat W_i&=\mat I_K-\mat A_i\mat P_i\mat A_i^H.
\label{eq:jes_iteration_matrices}
\end{align}
Using the CIR parameterization \eqref{eq:ci_region}, the optimization of {$\widetilde{\mat S}$} in $\mathcal P_4$ separates into $T$ independent convex nonnegative least-squares (NNLS) problems \cite{lawson1995solving_ls}:
\begin{equation}
\boldsymbol\delta_i[t]
=\operatorname*{arg\,min}_{\boldsymbol\delta\succeq\vect0}
\left\|\mat L_i\bigl(\bar{\vect s}[t]
+\boldsymbol\Lambda[t]\boldsymbol\delta\bigr)\right\|_2^2,
\label{eq:target_nnls}
\end{equation}
where $\mat L_i$ is a square-root factor of the real representation of $\mat W_i$:
\begin{equation}
\overline{\mat W}_i=
\begin{bmatrix}
\Re\{\mat W_i\}&-\Im\{\mat W_i\}\\
\Im\{\mat W_i\}& \Re\{\mat W_i\}
\end{bmatrix},
\qquad
\mat L_i^T\mat L_i=\overline{\mat W}_i.
\label{eq:real_ci_weight}
\end{equation}
Since $\xi>0$, $\mat W_i$ is positive definite. Substituting the optimized CIR coefficients $\boldsymbol\delta_i[t]$ into \eqref{eq:ci_region} and converting the real representation back to complex form gives the target signal block $\widetilde{\mat S}_i$. Substituting $\FM(\mat A_i)$ from \eqref{eq:metric_realization} into \eqref{eq:normalized_rf_update} then yields:
\begin{equation}
\mat V_i=\mat D_i^{1/2}\mat P_i\mat A_i^H\widetilde{\mat S}_i.
\label{eq:jes_digital_update}
\end{equation}
}

\subsubsection{Effective Channel Subspace Update}
{With the current CIR target block $\widetilde{\mat S}_i$ fixed, set $\mat A_{i+1}$ to $\mat A^\star$ in \eqref{eq:exact_subspace_update} according to \propref{prop:exact_target}.}

\begin{algorithm}[t]
\caption{JES-MiLAC-SLP Algorithm}
\label{alg:jes}
\begin{algorithmic}[1]
\STATE \textbf{Input:} $\mat H$, $\mat S$, $R$, $P_{\rm T}$, $\sigma^2$, $I_{\rm J}$, and $\epsilon_{\rm rel,J}$.
\STATE $\xi=K\sigma^2/P_{\rm T}$, $\mat G_H=(\mat H\mat H^H)^{-1}$, and $\mat C_0=\mat S\mat S^H/T$.
\STATE {$\mat H=\mat L_H\boldsymbol\Sigma_H\mat J_H^H$; compute $\mat B_\xi$ from \eqref{eq:fixed_target_weighted_matrix}.}
\STATE $\mat C_0=\mat E_0\operatorname{diag}(\nu_1,\ldots,\nu_K)\mat E_0^H$, \ $\nu_1\geq\cdots\geq\nu_K$.
\STATE $\mat A_0=[\vect e_{0,1},\ldots,\vect e_{0,R}]$.
\STATE $\mat D_0=\mat A_0^H\mat G_H\mat A_0$, \ $\mat P_0=(\mat I_R+\xi\mat D_0)^{-1}$.
\STATE At $\mat A_0$, solve \eqref{eq:target_nnls} for $t=1,\ldots,T$.
\STATE {Obtain $\widetilde{\mat S}_0$ and $\mat V_0$ from \eqref{eq:ci_region} and \eqref{eq:jes_digital_update}.} Set $i=0$.
\STATE \textbf{repeat}
\STATE \quad {Compute $\mat M_i$ and $\mat E_{i,R}$ using \eqref{eq:fixed_target_weighted_matrix} and \eqref{eq:target_subspace_basis}.}
\STATE \quad {Update $\mat A_{i+1}$ using \eqref{eq:exact_subspace_update}.}
\STATE \quad $\mat D_{i+1}=\mat A_{i+1}^H\mat G_H\mat A_{i+1}$, \ $\mat P_{i+1}=(\mat I_R+\xi\mat D_{i+1})^{-1}$.
\STATE \quad $\mat W_{i+1}=\mat I_K-\mat A_{i+1}\mat P_{i+1}\mat A_{i+1}^H$, \ $\mat L_{i+1}^T\mat L_{i+1}=\overline{\mat W}_{i+1}$.
\STATE \quad Solve \eqref{eq:target_nnls} for $t=1,\ldots,T$.
\STATE \quad {Compute $\widetilde{\mat S}_{i+1}$ and $\mat V_{i+1}$ from \eqref{eq:ci_region} and \eqref{eq:jes_digital_update}.}
\STATE \quad $i=i+1$.
\STATE \textbf{until} a convergence criterion is satisfied.
\STATE $\displaystyle\FM=\FM(\mat A_i)$ from \eqref{eq:metric_realization}.
\STATE $\displaystyle\gamma=\sqrt{\frac{P_{\rm T}T}{\|\mat V_i\|_F^2}}$, and $\mat U=\gamma\mat V_i$.
\STATE \textbf{Output:} $\FM$, $\mat U$, $\widetilde{\mat S}_i$, and $\gamma$.
\end{algorithmic}
\end{algorithm}

\subsubsection{Alternating Updates}
{\algref{alg:jes} summarizes the initialization and alternating updates. {The optimization of the SLP variables {$\widetilde{\mat S}$ and $\mat V$} and {the {basis-matrix update of $\mat A$} in \propref{prop:exact_target}} ensure a nonincreasing objective. Since the MSE is bounded below by zero, the objective values converge.}}
The complexity of \algref{alg:jes} mainly lies in channel preprocessing, computing $\widetilde{\mat S}_i$ through the $T$ NNLS problems in \eqref{eq:target_nnls}, and updating the effective channel basis. Computing $\mat G_H$, $\mat H^\dagger$, $\mat A_0$, {$\mat L_H$, and $\mat B_\xi$} costs $\mathcal O(NK^2+K^2T+K^3)$.
Let $\bar L$ denote the average number of active-set least-squares solves for each NNLS problem.
The dominant complexity of computing $\widetilde{\mat S}_i$ once is $\mathcal O(T\bar L K^3)$.
{
Each subspace update costs at most $\mathcal O(K^2T+K^3)$, including forming $\mat M_i$, computing $\mat E_{i,R}$, and updating $\mat A_{i+1}$ via \eqref{eq:exact_subspace_update}. With $I_{\rm J}$ updates, the overall complexity is $\mathcal O\!\left(NK^2+(I_{\rm J}+1)
[K^2T+K^3+T\bar L K^3]\right).$
}

\begin{remark}
\label{rem:source_power}
The theoretical results and JES-MiLAC-SLP algorithm in this section also apply to the source-signal power constraint adopted in existing MiLAC studies \cite{wu2026milac_miso_limits}. For matched sources, normalizing the signals and power budget according to our signal definitions expresses this constraint as $\|\mat U\|_F^2/T\leq P_{\rm T}$. Under the column-orthonormal representation $\FM^H\FM=\mat I_R$ adopted in this section, $\|\FM\mat U\|_F^2=\|\mat U\|_F^2$. Hence, the RF-port input-power and antenna-port transmit-power constraints coincide, yielding the same optimum for the joint design. The theoretical results and algorithm in this section therefore remain applicable.
\end{remark}

\section{Hardware-Constrained MiLAC-SLP Design}
\label{sec:physical}
{The design in Section~\ref{sec:lossless} admits a stem-connected realization under the continuous-susceptance construction conditions, but the required susceptances may violate hardware range or quantization constraints. We therefore study joint MiLAC-SLP design under these practical constraints.}

\vspace{-2mm}
\subsection{Hardware-Constrained Stem-Connected MiLAC Model}
\label{subsec:stem_model}

We adopt the stem-connected topology and finite-range quantized susceptance model defined in Section~\ref{subsec:architecture_signal_model}. For a circuit state $\vect b$ satisfying \eqref{eq:quantized_susceptance}, we construct $\mat B_{\rm net}(\vect b)$ using \eqref{eq:network_susceptance}, obtain $\boldsymbol\Theta(\vect b)$ from \eqref{eq:cayley_network}, and take its lower-left block as $\FM(\vect b)$ according to \eqref{eq:milac_signal_mapping}. Under the same antenna-port power constraint, jointly optimizing the circuit state {$\vect b$}, RF-port input signals {$\mat U$}, target signals within the CIR {$\widetilde{\mat S}$}, and receive scaling factor {$\gamma$} gives
\begin{subequations}
\label{eq:stem_joint_problem}
\begin{align}
\mathcal P_5:\ 
\min_{\vect b,\mat U,\widetilde{\mat S},\gamma>0}\quad
&\frac{1}{KT}\mathbb E_{\mat N}\left\|
\frac{\mat H\FM(\vect b)\mat U+\mat N}{\gamma}-\widetilde{\mat S}
\right\|_F^2\\
\text{s.t.}\quad
&b_i\in\mathcal Q_q,\quad i=1,\ldots,N_{\rm c},\\
&\frac{1}{T}\|\FM(\vect b)\mat U\|_F^2\le P_{\rm T},
\quad \widetilde{\mat S}\in\mathcal D(\mat S).
\end{align}
\end{subequations}

{Problem $\mathcal P_5$ is a nonconvex discrete-continuous joint optimization problem: each branch susceptance takes values from a finite quantized alphabet, forming the discrete circuit state $\vect b$, which determines $\FM(\vect b)$ nonlinearly and is coupled with the continuous SLP variables. Moreover, the number of feasible circuit states grows exponentially with the number of tunable branches.}

\subsection{Fixed-Circuit Representation and Performance}
\label{subsec:physical_theory}

To evaluate a circuit state in HC-JES-MiLAC-SLP, we fix $\vect b$ in $\mathcal P_5$ and minimize over the continuous SLP variables $\mat U$, $\widetilde{\mat S}$, and $\gamma$. Unlike in Section~\ref{sec:lossless}, the actual analog beamforming matrix $\FM(\vect b)$ is generally not column orthonormal. We therefore represent it in column-orthonormal form and transform the scaled RF-domain signal variable accordingly, converting the fixed-circuit SLP subproblem into the same form as in Section~\ref{sec:lossless} and thereby enabling evaluation of the SLP-optimized performance for a fixed $\vect b$. For a retained state satisfying $\rank(\FM(\vect b))=R$, define
\begin{align}
\mat R_{\rm M}(\vect b)
&\triangleq
\left(\FM(\vect b)^H\FM(\vect b)\right)^{1/2},\\
\bar{\mat F}_{\rm M}(\vect b)
&\triangleq
\FM(\vect b)\mat R_{\rm M}(\vect b)^{-1}.
\label{eq:physical_polar_equivalence}
\end{align}
{The column-orthonormal matrix $\bar{\mat F}_{\rm M}(\vect b)$ is determined by $\vect b$. Correspondingly, redefining the scaled RF-domain coefficient block as $\mat V=\mat R_{\rm M}(\vect b)\mat U/\gamma$ gives}
\begin{equation}
\frac{\FM(\vect b)\mat U}{\gamma}
=\bar{\mat F}_{\rm M}(\vect b)\mat V,
\ \ 
\frac{1}{T}\left\|\frac{\FM(\vect b)\mat U}{\gamma}\right\|_F^2
=\frac{1}{T}\|\mat V\|_F^2.
\label{eq:physical_coordinate_change}
\end{equation}
Equivalently, the actual power of the original circuit is $\|\FM(\vect b)\mat U\|_F^2/T=\gamma^2\|\mat V\|_F^2/T$.
For a fixed $\bar{\mat F}$, the minimum objective value is given by
\begin{equation}
f(\bar{\mat F})
\triangleq
\min_{\mat V,\,\widetilde{\mat S}\in\mathcal D(\mat S)}
\frac{\|\mat H\bar{\mat F}\mat V-\widetilde{\mat S}\|_F^2}{KT}
+\frac{\sigma^2\|\mat V\|_F^2}{P_{\rm T}T}.
\label{eq:physical_fixed_subspace_value}
\end{equation}
Accordingly, the objective value of a circuit state is $\bar f(\vect b)=f(\bar{\mat F}_{\rm M}(\vect b))$.

The following proposition bounds the difference between the optimal MSE values corresponding to two column-orthonormal analog beamforming matrices.
\begin{proposition}[MiLAC Sensitivity]
\label{prop:stability}
Let $\widetilde{\mat S}_\ell^\star$ be an optimal CI target for $\bar{\mat F}_\ell$, and let $E_{\widetilde S}=\max\{\|\widetilde{\mat S}_1^\star\|_F^2,\|\widetilde{\mat S}_2^\star\|_F^2\}$. Then
\begin{equation}
\left|f(\bar{\mat F}_1)-f(\bar{\mat F}_2)\right|
\leq
\frac{\|\mat H\|_2^2E_{\widetilde S}}{KT\xi}
\left\|
\bar{\mat F}_1\bar{\mat F}_1^H-
\bar{\mat F}_2\bar{\mat F}_2^H
\right\|_2.
\label{eq:subspace_stability}
\end{equation}
\end{proposition}

\begin{proof}
See Appendix~\ref{app:proof_stability}.
\end{proof}

{This bound supports using the analog beamforming matrix returned by \algref{alg:jes} to construct initial circuit states.}

\vspace{-3mm}
\subsection{Algorithm Overview and Initialization}
\label{subsec:physical_algorithm}

Building on the preceding analysis, we propose HC-JES-MiLAC-SLP, summarized in \algref{alg:physical_jes}. The algorithm first initializes the circuit state by synthesizing and quantizing the JES-MiLAC-SLP solution as described below. It then alternates between optimizing $\mat U$, $\widetilde{\mat S}$, and $\gamma$ for fixed $\vect b$ using the procedure in this subsection and updating the quantized circuit state as detailed in Section~\ref{subsec:quantized_circuit_update}. 

\subsubsection{Circuit-State Initialization}
\label{subsubsec:circuit_initialization}

We use the closed-form synthesis method in \cite{nerini2026reduced_complexity_milac,zhang2026stem_connected_milac} to convert a given column-orthonormal target $\mat F$ into continuous branch susceptances, denoted by
\begin{equation}
\widetilde{\vect b}=\mathcal S_{\rm stem}(\mat F).
\label{eq:stem_synthesis_map}
\end{equation}
Let $\Pi_{\mathcal Q_q}$ denote the componentwise nearest-neighbor quantization map onto $\mathcal Q_q$. In particular,
\begin{equation}
[\Pi_{\mathcal Q_q}(\widetilde{\vect b})]_j\in
\operatorname*{arg\,min}_{q\in\mathcal Q_q}|\widetilde b_j-q|.
\label{eq:componentwise_quantizer}
\end{equation}
Denote $\bar{\mat F}_{\rm ideal}$ as the column-orthonormal analog beamforming matrix returned by \algref{alg:jes}.
{Right-unitary rotations preserve the ideal transmit block since $(\bar{\mat F}_{\rm ideal}\boldsymbol\Omega)(\boldsymbol\Omega^H\mat V)=\bar{\mat F}_{\rm ideal}\mat V$, but generally yield different circuit states after synthesis and quantization. We therefore generate multiple {rotated target matrices $\bar{\mat F}_{\rm ideal}\boldsymbol\Omega$} for initialization.}
Specifically, we form $\boldsymbol\Omega_0=\mat I_R$ and $\boldsymbol\Omega_m=\operatorname{qf}(\mat Z_m)$ for $m=1,\ldots,M_{\rm orb}-1$, where $\mat Z_m$ has independent circular Gaussian entries and $\operatorname{qf}(\mat Z_m)$ denotes the unitary factor in its thin QR decomposition.
Let $\mathcal M_{\rm syn}$ contain the indices for which $\mathcal S_{\rm stem}(\bar{\mat F}_{\rm ideal}\boldsymbol\Omega_m)$ succeeds. 
The circuit candidates are
\begin{equation}
\vect b_m=
\Pi_{\mathcal Q_q}\!\left(
\mathcal S_{\rm stem}(\bar{\mat F}_{\rm ideal}\boldsymbol\Omega_m)
\right),
\ \  m\in\mathcal M_{\rm syn}.
\label{eq:orbit_candidates}
\end{equation}
{After quantization, we retain only candidates with full-column-rank analog beamforming matrices {$\FM(\vect b_m)$}, optimize the SLP variables {$\mat U$, $\widetilde{\mat S}$, and $\gamma$} for each, and select the state with the smallest $\bar f(\vect b_m)$ for initialization. Thus,}
{
\begin{equation}
m^\star\in\argmin_{\substack{m\in\mathcal M_{\rm syn}\\
\rank(\FM(\vect b_m))=R}}
\bar f(\vect b_m),
\qquad
\vect b^{(0)}=\vect b_{m^\star}.
\label{eq:orbit_initialization}
\end{equation}
}
Initialization requires at least one synthesized candidate to remain full column rank after quantization.
{As an optional enhancement, run $I_{\rm pre}$ HC-JES-MiLAC-SLP updates from each of the $N_{\rm keep}$ states with the smallest initial $\bar f(\vect b_m)$, then continue from the resulting state with the smallest $\bar f(\vect b)$.}

\subsubsection{Fixed-Circuit SLP Update}
\label{subsubsec:fixed_circuit_slp_update}

{At iteration $i$, fix $\vect b^{(i)}$ and set $\bar{\mat F}_i=\bar{\mat F}_{\rm M}(\vect b^{(i)})$. Obtain the weight matrix from \eqref{eq:reduced_mse_weight} with $\FM=\bar{\mat F}_i$, and solve \eqref{eq:target_nnls} for the CIR targets {$\widetilde{\mat S}^{(i)}$}. Compute $\mat V^{(i)}$ from \eqref{eq:normalized_rf_update} using the same substitution, and recover $\gamma^{(i)}$ from \eqref{eq:ci_scaling_recovery} and $\mat U^{(i)}=\gamma^{(i)}\mat R_{\rm M}(\vect b^{(i)})^{-1}\mat V^{(i)}$.}

\begin{algorithm}[t]
\caption{HC-JES-MiLAC-SLP Algorithm}
\label{alg:physical_jes}
\begin{algorithmic}[1]
\STATE \textbf{Input:} $\mat H$, $\mat S$, $R$, $P_{\rm T}$, $\sigma^2$, stem topology, $\mathcal Q_q$, $M_{\rm orb}$, $\Delta_{\max}$, $\eta$, $I_{\rm P}$, and $\epsilon_{\rm rel}$.
\STATE Obtain $\bar{\mat F}_{\rm ideal}$ from \algref{alg:jes}; construct $\vect b_m$ by \eqref{eq:orbit_candidates}.
\STATE Retain full-column-rank candidates and select $\vect b^{(0)}$ by \eqref{eq:orbit_initialization}. Set $i=0$.
\STATE \textbf{repeat}
\STATE \quad Solve the fixed-circuit SLP problem at $\vect b^{(i)}$; obtain $\widetilde{\mat S}^{(i)}$, $\mat U^{(i)}$, $\gamma^{(i)}$, and $\bar f(\vect b^{(i)})$.
\STATE \quad {Compute $\bar{\mat F}_{\rm M}^{(i)}$ and $\mat V^{(i)}$ using \eqref{eq:physical_polar_equivalence} and \eqref{eq:physical_coordinate_change}.}
{\STATE \quad {Form $\vect r^{(i)}$ and $\mat J^{(i)}$ by \eqref{eq:physical_projected_model}.}}
\STATE \quad Compute $\vect d_{\rm b}^{(i)}$ from \eqref{eq:physical_direction}.
\STATE \quad Construct $\mathcal C_i$ by \eqref{eq:quantized_ray_points}--\eqref{eq:quantized_ray_candidates}.
\STATE \quad Re-solve the fixed-circuit SLP problem for all $\vect b\in\mathcal C_i$; accept $\vect b_\star^{(i)}$ only when the relative decrease exceeds $\epsilon_{\rm rel}$. Otherwise, stop.
\STATE \quad $i=i+1$.
\STATE \textbf{until} a convergence criterion is satisfied.
\STATE \textbf{Output:} $\vect b$, $\FM(\vect b)$, $\mat U$, $\widetilde{\mat S}$, and $\gamma$.
\end{algorithmic}
\end{algorithm}

\vspace{-2mm}
\subsection{Quantized Circuit-State Update}
\label{subsec:quantized_circuit_update}
{To optimize the discrete circuit state $\vect b$, we construct a first-order Taylor linearization of the residual vector at the current iterate} while accounting for adjustable changes in the SLP variables {$\mat V$ and $\boldsymbol\delta$}. This yields a continuous update direction for $\vect b$ and guides the design of the quantized circuit-state update.

\subsubsection{Local Continuous Update}

We first construct a real residual vector $\vect e^{(i)}\in\mathbb R^{2(K+N)T}$ such that
\begin{equation}
\bar f(\vect b^{(i)})={\|\vect e^{(i)}\|_2^2}/{KT}.
\label{eq:physical_local_residual}
\end{equation}
{
Specifically,
\begin{equation}
\vect e^{(i)}=
\begin{bmatrix}
\mathcal R\!\left(\mat H\bar{\mat F}_{\rm M}^{(i)}\mat V^{(i)}
-\widetilde{\mat S}^{(i)}\right)\\
\sqrt{\xi}\,\mathcal R\!\left(\bar{\mat F}_{\rm M}^{(i)}\mat V^{(i)}\right)
\end{bmatrix}
\in\mathbb R^{2(K+N)T}.
\label{eq:physical_residual_vector}
\end{equation}
where $\mathcal R(\mat A)$ stacks the vectorized real and imaginary parts of a complex matrix $\mat A$ into a real vector.
}

{
{Using the coordinate change in \eqref{eq:physical_coordinate_change}, the CIR parameterization in \eqref{eq:ci_region}, and elimination of the receive scaling factor, problem \eqref{eq:stem_joint_problem} depends only on $\mat V$, $\boldsymbol\delta$, and $\vect b$. We jointly linearize the residual with respect to these variables to determine the circuit update direction.}
Specifically, let $\vect d_{\rm V}\in\mathbb R^{2RT}$ collect the reshaped real representation of the changes of $\mat V$, and let $\vect d_{\rm b}\in\mathbb R^{N_{\rm c}}$ represent the change of ${\bf b}$.
By \eqref{eq:ci_region}, the CIR coefficients $\boldsymbol\delta$ must be nonnegative. We therefore vary only the currently positive coefficients and define their index set as}
\begin{equation}
\mathcal A_i
\triangleq
\{(t,\ell):\delta_\ell^{(i)}[t]>0\}.
\label{eq:active_ci_offset_set}
\end{equation}
Based on this, we further define $\vect d_\delta\in\mathbb R^{|\mathcal A_i|}$ as the changes of $\boldsymbol{\delta}^{(i)}[t]$ indexed by $\mathcal A_i$.
{The first-order Taylor linearization of the residual vector at the current iterate is \cite{nocedal2006numerical}}
{
\begin{equation}
\vect e_{\rm loc}^{(i)}(\vect d_{\rm V},\vect d_\delta,\vect d_{\rm b})
\triangleq\vect e^{(i)}
+\mat J_{\rm V}^{(i)}\vect d_{\rm V}
+\mat J_\delta^{(i)}\vect d_\delta
+\mat J_{\rm b}^{(i)}\vect d_{\rm b},
\label{eq:physical_local_model}
\end{equation}
where the Jacobian matrices $\mat J_{\rm V}^{(i)}\in\mathbb R^{2(K+N)T\times2RT}$, $\mat J_\delta^{(i)}\in\mathbb R^{2(K+N)T\times|\mathcal A_i|}$, and $\mat J_{\rm b}^{(i)}\in\mathbb R^{2(K+N)T\times N_{\rm c}}$ contain the first-order partial derivatives of the residual vector with respect to $\vect d_{\rm V}$, $\vect d_\delta$, and $\vect d_{\rm b}$, respectively, at the current iterate.}
{The Jacobians are obtained by differentiating the residual expression with respect to the corresponding variables. Explicit expressions are omitted for brevity.}
To eliminate the two SLP-variable updates {$\vect d_{\rm V}$ and $\vect d_\delta$}, we first group their derivative matrices and define the residual and circuit-state derivative remaining after orthogonal projection:
{
\begin{equation}
\begin{gathered}
\mat L^{(i)}=[\mat J_{\rm V}^{(i)},\mat J_\delta^{(i)}], \quad\boldsymbol\Pi^{(i)}=\mat I-\mat L^{(i)}(\mat L^{(i)})^\dagger,\\
\vect r^{(i)}=\boldsymbol\Pi^{(i)}\vect e^{(i)},\quad
\mat J^{(i)}=\boldsymbol\Pi^{(i)}\mat J_{\rm b}^{(i)}.
\end{gathered}
\label{eq:physical_projected_model}
\end{equation}
}
For a fixed $\vect d_{\rm b}$, minimizing the linearized residual over $\vect d_{\rm V}$ and $\vect d_\delta$ yields an expression that depends only on $\vect d_{\rm b}$:
{
\begin{equation}
\begin{aligned}
\min_{\vect d_{\rm V},\vect d_\delta}\,
&\|\vect e^{(i)}+\mat J_{\rm V}^{(i)}\vect d_{\rm V}+
\mat J_\delta^{(i)}\vect d_\delta+\mat J_{\rm b}^{(i)}\vect d_{\rm b}\|_2^2\\
=&\|\vect r^{(i)}+\mat J^{(i)}\vect d_{\rm b}\|_2^2.
\end{aligned}
\label{eq:physical_elimination}
\end{equation}
}

We minimize this objective function subject to the susceptance range and per-update change limits, with regularization to discourage excessively large continuous increments. 
Let $\Delta_{\max}>0$ be the maximum continuous susceptance change per update and define
\begin{align}
\boldsymbol\ell^{(i)}&=\max\{-B_{\max}\mat 1-\vect b^{(i)},-\Delta_{\max}\mat 1\},\\
\boldsymbol u^{(i)}&=\min\{B_{\max}\mat 1-\vect b^{(i)},\Delta_{\max}\mat 1\},
\label{eq:physical_direction_bounds}
\end{align}
where $\mat1\in\mathbb R^{N_{\rm c}}$ is the all-ones vector, and $\min$ and $\max$ act elementwise.
The continuous susceptance increment is obtained by solving the following regularized least-squares problem with componentwise bounds:
\begin{align}
\vect d_{\rm b}^{(i)}\in{}&\argmin_{\boldsymbol\ell^{(i)}\preceq\vect d_{\rm b}\preceq\boldsymbol u^{(i)}}
\|\vect r^{(i)}+\mat J^{(i)}\vect d_{\rm b}\|_2^2\nonumber\\
&\qquad+\eta\lambda_{\max}[(\mat J^{(i)})^T\mat J^{(i)}]\|\vect d_{\rm b}\|_2^2,
\label{eq:physical_direction}
\end{align}
where $\eta\geq0$ is the regularization coefficient{, and $\lambda_{\max}(\cdot)$ denotes the largest eigenvalue}.
{This convex quadratic subproblem can be solved using a bounded-variable least-squares method \cite{lawson1995solving_ls}.}

\subsubsection{Quantized Candidate Selection}

Starting from $\vect b^{(i)}$, we form all distinct quantized circuit states obtained by scaling the continuous update $\vect d_{\rm b}^{(i)}$ by a factor between zero and one. Let $\bar\tau_{i,\ell}\in[0,1]$ denote the factor for the $\ell$-th state. Then
\begin{equation}
\vect b_{i,\ell}=
\Pi_{\mathcal Q_q}(
\vect b^{(i)}+\bar\tau_{i,\ell}\vect d_{\rm b}^{(i)}
),
\label{eq:quantized_ray_points}
\end{equation}
Retaining only candidates satisfying $\rank(\FM(\vect b_{i,\ell}))=R$, together with the current state, gives
\begin{equation}
\mathcal C_i=\{\vect b^{(i)}\}\cup
\left\{\vect b_{i,\ell}:\rank(\FM(\vect b_{i,\ell}))=R\right\}.
\label{eq:quantized_ray_candidates}
\end{equation}
For each $\vect b\in\mathcal C_i$, reoptimize the SLP variables {$\mat U$, $\widetilde{\mat S}$, and $\gamma$} using the fixed-circuit procedure in Section~\ref{subsubsec:fixed_circuit_slp_update} and evaluate $\bar f(\vect b)$. Let $\textstyle\vect b_\star^{(i)}\in\argmin_{\vect b\in\mathcal C_i}\bar f(\vect b).$ $\vect b_\star^{(i)}$ is accepted for the update if it provides a sufficient relative objective improvement.

\vspace{-2mm}
\subsection{Complexity and Extensions}
\label{subsec:physical_complexity}

The main computations in \algref{alg:physical_jes} are obtaining the susceptance increment $\vect d_{\rm b}^{(i)}$ from \eqref{eq:physical_direction} and reoptimizing SLP for each candidate circuit state in $\mathcal C_i$. Each candidate requires computing the analog beamforming matrix, the effective channel and associated matrices, and solving $T$ NNLS problems. Let $I_{\rm H}$ denote the total number of circuit updates including initialization screening, $\bar N_{\rm cand}$ the average number of candidates evaluated per update, and $\bar L$ the average number of active-set least-squares solves per NNLS problem. When candidate evaluation dominates the computation, the dominant complexity is $\mathcal O(I_{\rm H}\bar N_{\rm cand}[(N+R)^3+KNR+K^2T+T\bar L K^3])$.
{Each accepted circuit update strictly decreases $\bar f(\vect b)$, evaluated after reoptimizing the SLP variables. Since the feasible discrete circuit set is finite, HC-JES-MiLAC-SLP terminates after finitely many updates.}

\begin{remark}
\label{rem:hc_rf_input_power}
\algref{alg:physical_jes} also applies to the RF-port input-signal power constraint $\|\mat U\|_F^2/T\leq P_{\rm T}$ with changes only to the fixed-circuit performance evaluation and circuit-update direction. For a fixed $\vect b$, redefine $\mat V=\mat U/\gamma$ and minimize \eqref{eq:physical_fixed_subspace_value} with $\bar{\mat F}$ replaced by $\FM(\vect b)$. The target signal block {$\widetilde{\mat S}$} is still optimized by eliminating the digital variable {$\mat V$} and solving NNLS problems. To update the circuit state, we reconstruct the first-order residual model from this objective, $\FM(\vect b)$, and its derivatives, and then compute the susceptance increment {$\vect d_{\rm b}$}.
\end{remark}

\vspace{-2mm}
\section{Numerical Results}
\label{sec:numerical}
\subsection{Simulation Configuration}
\label{subsec:simulation_configuration}

\begin{table}[t]
\caption{Simulation parameters \cite{wang2022ci_wmmse,wang2024average_ser_slp,fang2026lossless_reciprocal_milac,wu2026milac_miso_limits,zhang2026stem_connected_milac}.}
\label{tab:simulation_parameters}
\centering
\scriptsize
\setlength{\tabcolsep}{3pt}
\begin{tabular}{p{0.50\columnwidth}p{0.40\columnwidth}}
\toprule
Parameter & Value \\
\midrule
Users and antennas & $K=8$, $N\in\{14, 16^\dagger, 20, 24\}$ \\
RF-chain number & $R\in\{6^\dagger, 7, 8\}$ \\
Symbol-block length & $T=10$ \\
Modulation constellation & 16QAM \\
Channel model & 3GPP\_UMa\_NLOS \\
Power budget & $P_{\rm T}=1$ \\
SNR range & $0,5,\ldots,30$ dB \\
Stem reference admittance & $Y_0=0.02$ S \\
Susceptance range & $B_{\max}=0.007$ S \\
Quantization resolution & $q\in\{2,3^\dagger,4\}$ bits \\
Number of adjustable susceptances & $N_{\rm c}=R(2N+1)$ \\
Maximum outer iterations & 50 \\
Right-unitary initial representations & $M_{\rm orb}=8$ \\
Maximum susceptance change & $\Delta_{\max}=4\Delta_q$ \\
Regularization coefficient & $\eta=10^{-4}$ \\
Relative objective tolerances & $10^{-5}$ (JES), $10^{-6}$ (HC-JES)\\
Monte Carlo realizations & $300$ \\
\bottomrule
\end{tabular}
\parbox{0.94\columnwidth}{\scriptsize $\dagger$ Default value, with variations specified for the relevant simulations.}
\vspace{-4mm}
\end{table}
 
In this section, we employ the Monte Carlo method to assess the performance of the proposed schemes. 
{The simulation SNR is defined as $P_{\rm T}/\sigma^2$.}
Unless otherwise stated, all channels are generated by QuaDRiGa \cite{jaeckel2014quadriga} under the {urban macrocell (UMa) non-line-of-sight (NLoS)} scenario specified in 3GPP TR 38.901 \cite{3gpp_tr_38_901}. The {base station} employs a horizontal {uniform linear array (ULA)} based on the ``3gpp-3d'' antenna model, each user has a single omnidirectional antenna, and the carrier frequency is 3.5 GHz. The users are randomly distributed over a $120^\circ$ sector with radii from 35 to 500 m. 
For ease of comparison, we consider normalized channels satisfying $\mathbb E\{\operatorname{tr}(\mat H\mat H^H)\}= KN$.
Although unit-average-power QAM is employed, the proposed methods are not restricted to the particular modulation order adopted in Table~\ref{tab:simulation_parameters}. For higher-order QAM or PSK, only the CIR boundary matrix $\boldsymbol\Lambda[t]$ needs to be adjusted according to the corresponding constellation \cite{wang2024robust_aging_slp}. 
All parameter points are averaged over independently generated $(\mat H,\mat S)$ samples.
The reported MSE is the objective in {\eqref{eq:lr_joint_objective}}, and SER is computed from the exact closed-form of SER \cite{wang2024average_ser_slp}.
The stem-connected MiLAC follows the topology in \cite{nerini2026reduced_complexity_milac,zhang2026stem_connected_milac}, with its susceptances uniformly quantized according to \eqref{eq:quantized_susceptance}. Except for the parameter being swept, the default values in Table~\ref{tab:simulation_parameters} are used.
All methods are compared under the same power constraint and using the same $(\mat H,\mat S)$ samples. The methods in the figures are defined as follows.
\begin{itemize}
\item \textbf{FD-SLP}: fully digital SLP obtained by minimizing the same MSE objective \cite{wang2022ci_wmmse}.
\item \textbf{CB-MiLAC-SLP}: CB denotes channel-based MiLAC design. Its $\FM$ is formed from the $R$ dominant right singular vectors of $\mat H$ to retain the maximum channel energy \cite{wu2026milac_miso_limits}, after which the SLP variables are optimized.
\item \textbf{DHC-CB-MiLAC-SLP}: DHC denotes direct hardware-constrained mapping. \textbf{CB-MiLAC-SLP} is synthesized by the stem network \cite{zhang2026stem_connected_milac} and quantized to the available susceptance alphabet, after which the SLP variables are further optimized.
\item \textbf{Fully-Connected Hybrid SLP}: SLP under an ideal fully connected hybrid architecture with continuous phase shifters, where the block-fixed analog precoder and SLP are jointly optimized by {alternating optimization} \cite{lee2023_hybrid_mse_slp,lin2019_hybrid_mmse_mo,wang2022ci_wmmse}.
\item \textbf{JES-MiLAC-SLP}: \algref{alg:jes}, which jointly optimizes the effective channel subspace and the SLP variables under the general lossless reciprocal MiLAC model.
\item \textbf{DHC-JES-MiLAC-SLP}: The JES-MiLAC-SLP is directly mapped to the hardware via \eqref{eq:orbit_candidates} and selecting the candidate with the lowest optimized MSE via \eqref{eq:orbit_initialization}. 
\item \textbf{HC-JES-MiLAC-SLP}: \algref{alg:physical_jes}, which jointly optimizes the circuit state and the SLP variables under the hardware-constrained stem-connected topology and quantized susceptance alphabet.
\end{itemize}
\subsection{MiLAC SLP Performance}
\label{subsec:physical_layer_results}

\begin{figure}[!t]
\centering
\subfigure[SER versus SNR.] {\includegraphics[width=0.49\linewidth]{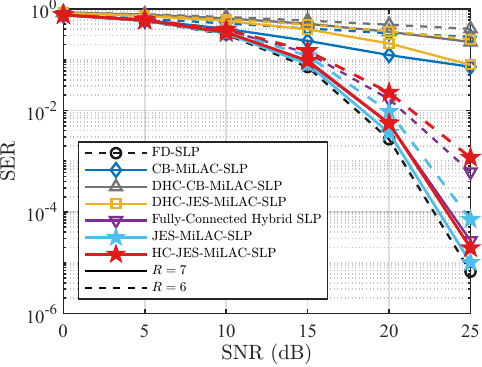}\label{fig:physical_snr_ser}}
\subfigure[MSE versus SNR.] {\includegraphics[width=0.49\linewidth]{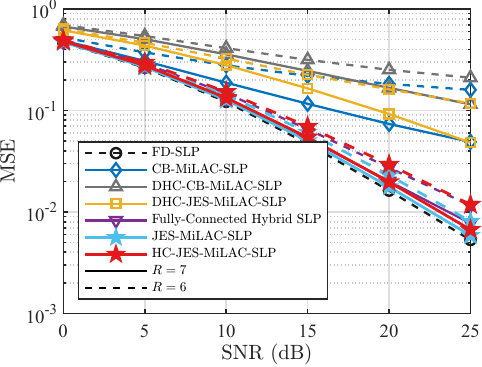}\label{fig:physical_snr_mse}}
\caption{Average SER and MSE versus SNR.}
\label{fig:physical_snr}
\vspace{-4mm}
\end{figure}

{\figref{fig:physical_snr} compares the SER and MSE of the seven methods for $R=6$ and $R=7$, highlighting the gains from symbol-aware subspace design, JES-based circuit initialization, and quantized circuit-state optimization.}

For both RF-chain configurations, the MSE and SER of CB-MiLAC-SLP approach high-SNR floors, whereas JES-MiLAC-SLP continues to improve with SNR and approaches FD-SLP. This gap is more pronounced for $R=6$. CB-MiLAC-SLP retains the dominant channel directions without accounting for the target signal that may be selected within the CIR for the current symbol block. \propref{prop:tail} shows that, when $R<K$, the high-SNR residual depends on the match between the effective channel subspace and this target signal block. This explains why JES-MiLAC-SLP markedly lowers the error floor of the channel-based design.
Under the same stem topology and susceptance alphabet, DHC-JES-MiLAC-SLP outperforms DHC-CB-MiLAC-SLP for both RF-chain configurations. Hence, the effective channel subspace designed for the current symbol block still yields a better initial circuit state after stem synthesis and quantization. HC-JES-MiLAC-SLP then optimizes the quantized circuit state and further reduces both MSE and SER. Its gap from JES-MiLAC-SLP reflects the restrictions that the stem topology, finite susceptance range, and finite-resolution quantization impose on the realizable analog beamforming matrices.
Compared with Fully-Connected Hybrid SLP, JES-MiLAC-SLP achieves lower MSE and SER for both RF-chain configurations over the entire SNR range, demonstrating the greater matrix-design flexibility of the general lossless reciprocal MiLAC over constant-modulus hybrid precoding. Despite the stem topology and quantized susceptance constraints, HC-JES-MiLAC-SLP achieves performance close to Fully-Connected Hybrid SLP over the SNR range. According to {\eqref{eq:lr_joint_objective}}, the noise term at low SNR weakens the differences among the analog beamforming matrices in signal fitting, whereas the benefits of designing the analog beamforming matrix and optimizing the quantized circuit state become more pronounced at high SNR.

\begin{figure}[!t]
\centering
\subfigure[JES-MiLAC-SLP.] {\includegraphics[width=0.49\linewidth]{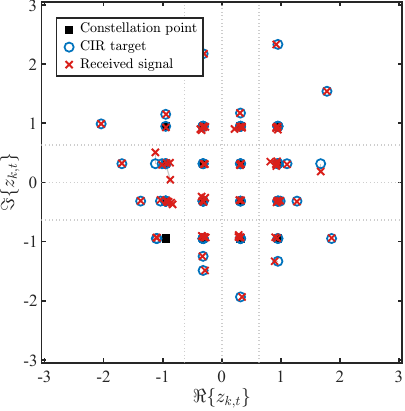}\label{fig:received_jes}}
\subfigure[HC-JES-MiLAC-SLP.] {\includegraphics[width=0.49\linewidth]{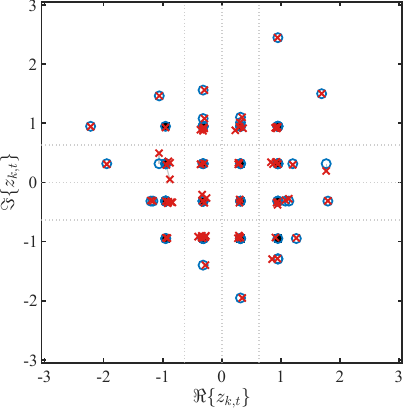}\label{fig:received_hc_jes}}
\caption{Received-signal distributions optimized at an SNR of 20 dB, plotted without added noise to show the deviations between the received and target signals.}
\label{fig:received_signal_snapshot}
\vspace{-3mm}
\end{figure}

{\figref{fig:received_signal_snapshot} shows the noiseless scaled signals $z_{k,t}=[\mat H\FM\mat U]_{k,t}/\gamma$ and CIR targets {of JES-MiLAC-SLP and HC-JES-MiLAC-SLP} for the same channel and symbol block, optimized at 20 dB. {Each method uses its own optimized receive scaling factor. Noise power is included in the optimization, but noise is omitted from the plot to isolate the fitting errors of the designs. Gray segments connect each received signal to its target, with shorter segments indicating smaller mismatch.} Both methods adjust the targets within the CIR while keeping the received signals close to them. Hardware constraints increase some HC-JES-MiLAC-SLP residuals, although most signals remain within their decision regions.}

\subsection{Configuration Effects}
\label{subsec:configuration_effects}

\begin{figure}[!t]
\centering
\subfigure[SER versus $N$.] {\includegraphics[width=0.49\linewidth]{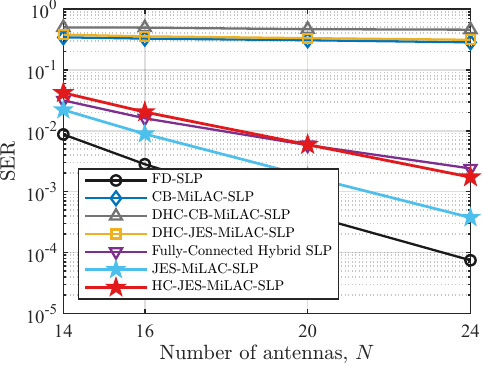}\label{fig:antenna_ser}}
\subfigure[MSE versus $N$.] {\includegraphics[width=0.49\linewidth]{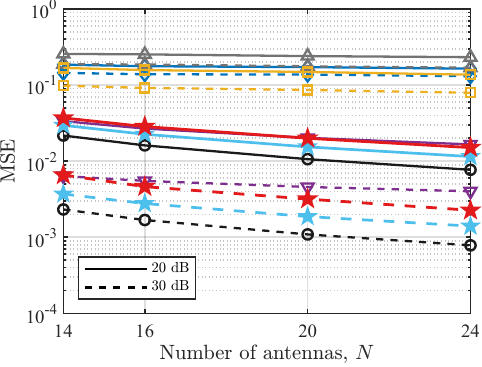}\label{fig:antenna_mse}}
\caption{Average SER and MSE versus the antenna number.}
\label{fig:physical_antenna_effect}
\vspace{-3mm}
\end{figure}

\begin{figure}[!t]
\centering
\subfigure[SER versus $R$.] {\includegraphics[width=0.49\linewidth]{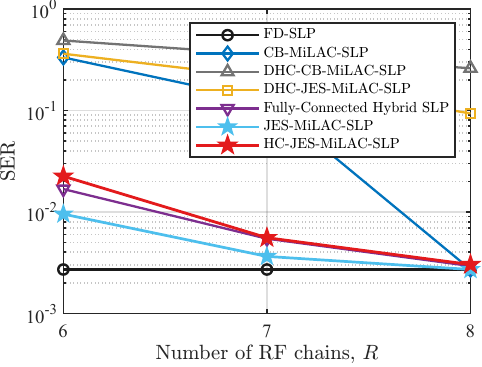}\label{fig:rf_ser}}
\subfigure[MSE versus $R$.] {\includegraphics[width=0.49\linewidth]{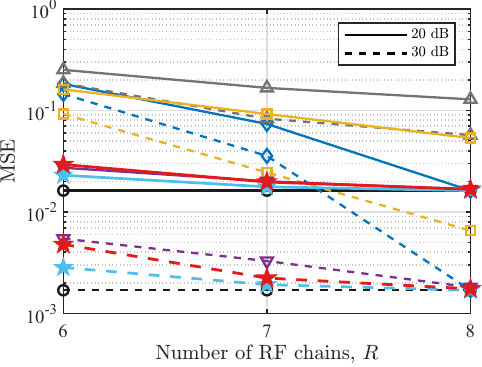}\label{fig:rf_mse}}
\caption{Average SER and MSE versus the RF chain number.}
\label{fig:physical_rf_effect}
\vspace{-2mm}
\end{figure}

\begin{figure}[!t]
\centering
\subfigure[SER versus $q$.] {\includegraphics[width=0.49\linewidth]{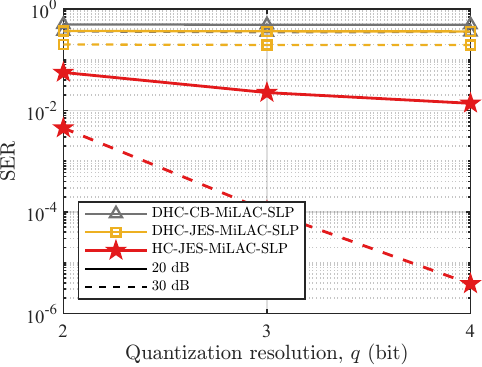}\label{fig:quantization_ser}}
\subfigure[MSE versus $q$.] {\includegraphics[width=0.49\linewidth]{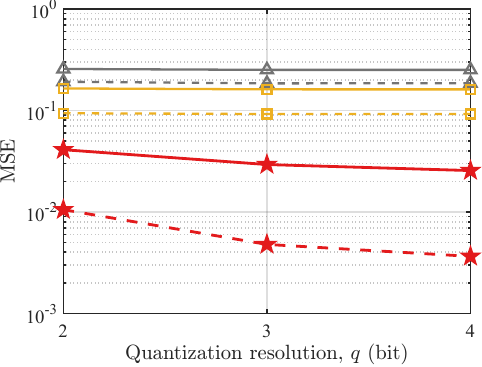}\label{fig:quantization_mse}}
\caption{Average SER and MSE versus the susceptance quantization resolution.}
\label{fig:physical_quantization_effect}
\vspace{-3mm}
\end{figure}

\figref{fig:physical_antenna_effect} illustrates the impact of the number of transmit antennas on the SER and MSE. As $N$ increases, the additional antenna degrees of freedom gradually improve the MSE and SER of all methods. Across the considered SNRs, HC-JES-MiLAC-SLP achieves performance close to or better than Fully-Connected Hybrid SLP and clearly outperforms the two DHC methods with fixed circuit states. At 30 dB, some SER values are very small and exhibit relatively large Monte Carlo fluctuations, and the corresponding curves are therefore omitted.
{\figref{fig:physical_rf_effect} reports SER at 20 dB and MSE at 20 and 30 dB as $R$ increases from $6$ to $K=8$. All constrained methods improve, and JES-MiLAC-SLP coincides with FD-SLP at $R=K$, confirming the lossless condition in Section~\ref{sec:lossless}. Fully-Connected Hybrid SLP and HC-JES-MiLAC-SLP also approach FD-SLP despite their hardware constraints, whereas the remaining DHC-JES-MiLAC-SLP gap highlights the need for circuit-state optimization.}
{\figref{fig:physical_quantization_effect} compares the three hardware-constrained MiLAC methods at 20 and 30 dB. Increasing $q$ reduces the quantization interval in \eqref{eq:quantization_spacing}, but yields only modest gains for the fixed-state DHC methods. In contrast, HC-JES-MiLAC-SLP markedly improves both SER and MSE, demonstrating the benefit of jointly optimizing the circuit state and SLP to exploit finer susceptance resolution.}

\begin{figure}[!t]
\centering
\subfigure[{MSE and its lower bound with fixed CIR targets.}]{\includegraphics[width=0.49\columnwidth]{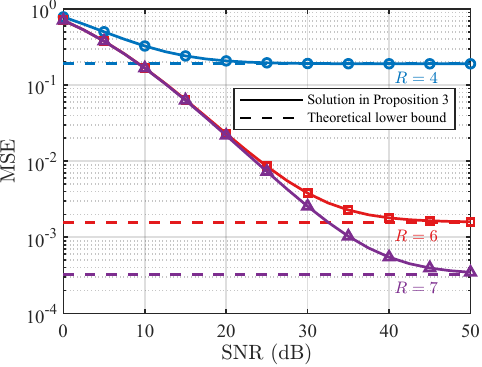}\label{fig:covariance_tail_bound}}
\hfill
\subfigure[{Convergence performance with $R=6$.}]{\includegraphics[width=0.49\columnwidth]{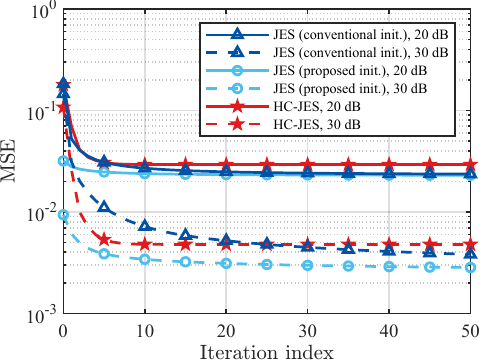}\label{fig:physical_convergence}}
\caption{{Verification of the theoretical lower bound and convergence performance.}}
\label{fig:convergence_and_bound}
\vspace{-3mm}
\end{figure}

{\figref{fig:covariance_tail_bound} compares the MSE attained by the solution in \propref{prop:exact_target} with the theoretical lower bound for different numbers of RF chains. With $K=8$ and $N=16$, the CIR targets are obtained by JES at $R=6$ and 20 dB and held fixed across all evaluated $R$ and SNR values. As the SNR increases, the MSE approaches the corresponding lower bound, confirming its asymptotic tightness for fixed CIR targets.}
{\figref{fig:physical_convergence} shows nonincreasing mean MSE objectives for both algorithms at $R=6$, consistent with their acceptance rules. JES-MiLAC-SLP improves progressively, whereas HC-JES-MiLAC-SLP achieves most of its reduction within the first few circuit updates.}
{We compare against conventional channel-SVD initialization with the $R$ dominant left singular vectors of $\mat H$. The proposed symbol-covariance initialization achieves a lower objective.}
\vspace{-3mm}
\section{Conclusion}
\label{sec:conclusion}
This work shows that the joint design of MiLAC and SLP can improve multiuser transmission by exploiting both channel state information and block-level user-symbol information. MiLAC connects large-scale antenna arrays with fewer RF chains, while SLP provides additional design freedom through target signals that exploit interference, thereby mitigating the performance loss caused by {a limited number of RF chains} and circuit constraints.
For ideal MiLAC networks, joint design preserves fully digital SLP performance when sufficient RF chains are available. With limited RF chains, performance depends on the match between the effective channel subspace selected by MiLAC and the selectable target signals. For practical stem-connected networks, the proposed hardware-constrained design narrows the performance gap to fully digital SLP under finite tuning ranges and discrete circuit states.

\appendices
\vspace{-2mm}
\section{Proof of Lemma~\ref{lem:representation}}
\label{app:proof_representation}
For any feasible $(\FM,\mat U)$ of \eqref{eq:lr_joint_problem}, define $\mat X=\FM\mat U$. Since $\FM\in\mathbb C^{N\times R}$, $\rank(\mat X)\leq R$. Choose any $\bar{\mat F}_{\rm M}\in\mathbb C^{N\times R}$ satisfying $\operatorname{Range}(\mat X)\subseteq\operatorname{Range}(\bar{\mat F}_{\rm M})$ and $\bar{\mat F}_{\rm M}^H\bar{\mat F}_{\rm M}=\mat I_R$, which exists because $N\geq R$. Set $\bar{\mat U}=\bar{\mat F}_{\rm M}^H\mat X$. Then $\bar{\mat F}_{\rm M}\bar{\mat U}=\bar{\mat F}_{\rm M}\bar{\mat F}_{\rm M}^H\mat X=\mat X=\FM\mat U$. Since $\bar{\mat F}_{\rm M}$ has unit spectral norm, the lossless reciprocal realization in \cite{wu2026milac_miso_limits} guarantees that it can be implemented by a feasible $\boldsymbol\Theta$. The original and constructed pairs therefore produce the same received signal, satisfy the same CI constraint, and use the same power. Thus, the two problems have the same optimum.
{This completes the proof.}
\vspace{-2mm}
\section{Proof of Lemma~\ref{prop:loss_regime}}
\label{app:proof_loss_regime}
For part (i), choose a semiunitary $\bar{\mat F}_{\rm M}\in\mathbb C^{N\times R}$ whose column space contains $\Range(\mat H^H)$, which exists since $K\leq R\leq N$. For any fully digital feasible block $\mat X$, set $\bar{\mat U}=\bar{\mat F}_{\rm M}^H\mat X$. Orthogonal projection gives $\mat H\bar{\mat F}_{\rm M}\bar{\mat U}=\mat H\mat X$ and $\|\bar{\mat F}_{\rm M}\bar{\mat U}\|_F\leq\|\mat X\|_F$.
This MiLAC realization is feasible by Lemma~\ref{lem:representation} and preserves the objective. Conversely, every feasible MiLAC transmit block is fully digital feasible, proving equality of the optima.
For part (ii), $\mat P_{\mat H\FM}\widetilde{\mat Y}=\widetilde{\mat Y}$, so orthogonal decomposition yields
\begin{align}
&{\|\widetilde{\mat Y}/\gamma-\widetilde{\mat S}\|_F^2}\notag\\
&=
{\|\widetilde{\mat Y}/\gamma-\mat P_{\mat H\FM}\widetilde{\mat S}\|_F^2}+
\|(\mat I_K-\mat P_{\mat H\FM})\widetilde{\mat S}\|_F^2.
\label{eq:projection_residual_decomposition}
\end{align}
Dropping the nonnegative noise term $\sigma^2/\gamma^2$ and the first term in \eqref{eq:projection_residual_decomposition}, then minimizing over the respective variables, gives \eqref{eq:rank_deficient_bound}.
{This completes the proof.}

\vspace{-2mm}
\section{Proof of Proposition~\ref{prop:metric_realization}}
\label{app:proof_metric_realization}
For a given feasible $\FM$, choose any $\mat A$ satisfying \eqref{eq:effective_range_inclusion}, which is possible because $\rank(\mat H\FM)\leq R\leq K$. Since $(\mat H^\dagger)^H\mat H^\dagger=\mat G_H$, \eqref{eq:metric_realization} gives $\FM(\mat A)^H\FM(\mat A)=\mat I_{R}$ and $\mat H\FM(\mat A)=\mat A\mat D_{\mat A}^{-1/2}$. Thus, $\FM(\mat A)$ is feasible. Because \eqref{eq:effective_range_inclusion} holds, there exists $\mat C\in\mathbb C^{R\times R}$, not necessarily nonsingular, such that $\mat H\FM=\mat A\mat C$. Write $\FM=\mat H^\dagger\mat A\mat C+\mat F_\perp$, where $\mat F_\perp=(\mat I_N-\mat H^\dagger\mat H)\FM$. The two terms are orthogonal and $\mat H\mat F_\perp=\mat0$. Since $\FM^H\FM=\mat I_R$, $\mat C^H\mat D_{\mat A}\mat C\preceq\mat I_R$. Therefore, $\mat C\mat C^H\preceq\mat D_{\mat A}^{-1}$ {hence}
\begin{align}
\mat H\FM\FM^H\mat H^H&=\mat A\mat C\mat C^H\mat A^H
\\
&\preceq\mat A\mat D_{\mat A}^{-1}\mat A^H
{=\mat H\FM(\mat A)\FM(\mat A)^H\mat H^H.}
\label{eq:dominance_gram}
\end{align}
For any $\mat X\in\mathbb C^{K\times R}$, define $\mat W_{\mat X}\triangleq\mat I_K-\mat X(\mat X^H\mat X+\xi\mat I_R)^{-1}\mat X^H$. Using $\mat X(\mat X^H\mat X+\xi\mat I)^{-1}\mat X^H=\mat X\mat X^H(\mat X\mat X^H+\xi\mat I)^{-1}$ and the operator monotonicity of $\mat X\mapsto\mat X(\mat X+\xi\mat I)^{-1}$ on positive semidefinite matrices \cite{horn2012matrix_analysis}, \eqref{eq:dominance_gram} yields $\mat W_{\mat H\FM(\mat A)}\preceq\mat W_{\mat H\FM}$. Hence, $f(\FM(\mat A))\leq f(\FM)$. Conversely, $\FM(\mat A)$ is feasible for every column-orthonormal $\mat A$.
Taking the minimum over the two feasible sets proves \eqref{eq:fm_a_value_equivalence}.
{This completes the proof.}

\vspace{-2mm}
\section{Proof of Proposition~\ref{prop:tail}}
\label{app:proof_tail}
{
Let $F_\xi(\mat A,\widetilde{\mat S})$ be the joint objective in \eqref{eq:subspace_value}, with
\begin{equation*}
m_\xi\triangleq\inf_{\mat A^H\mat A=\mat I_R,\,\widetilde{\mat S}\in\mathcal D(\mat S)}
F_\xi(\mat A,\widetilde{\mat S})=\widetilde f_\xi^\star,
\end{equation*}
{where $\inf$ denotes the infimum.}
Define $F_0$ by setting $\mat P_{\mat A}=\mat I_R$ and $m_0=\inf F_0$ over the same feasible set. Since $\mat P_{\mat A}\preceq\mat I_R$, $F_\xi\geq F_0$ and $m_\xi\geq m_0$.
For any $\epsilon>0$, choose a finite feasible pair satisfying $F_0(\mat A_\epsilon,\widetilde{\mat S}_\epsilon)\leq m_0+\epsilon$. At this fixed pair, $F_\xi\to F_0$, and hence
\begin{equation}
\begin{aligned}
m_0&\leq\liminf_{\xi\downarrow0}m_\xi
\leq\limsup_{\xi\downarrow0}m_\xi\\
&\leq F_0(\mat A_\epsilon,\widetilde{\mat S}_\epsilon)
\leq m_0+\epsilon,
\end{aligned}
\end{equation}
{where $\liminf$ and $\limsup$ denote the lower and upper limits, respectively.}
Letting $\epsilon\downarrow0$ gives $\lim_{\xi\downarrow0}m_\xi=m_0$. For fixed $\widetilde{\mat S}$, the Ky Fan principle \cite{horn2012matrix_analysis} yields
\begin{equation*}
\textstyle
\min_{\mat A^H\mat A=\mat I_R}F_0(\mat A,\widetilde{\mat S})
=\frac{1}{K}\sum_{i=R+1}^{K}\lambda_i\!\left(
{\widetilde{\mat S}\widetilde{\mat S}^H}/{T}\right).
\end{equation*}
Taking the infimum over $\widetilde{\mat S}\in\mathcal D(\mat S)$ yields \eqref{eq:tail_formula}.
{This completes the proof.}
}

{
  \vspace{-2mm}
\section{Proof of Proposition~\ref{prop:exact_target}}
\label{app:proof_exact_target}
Let $\mat C=\mat I_K+\xi(\mat H\mat H^H)^{-1}\succ0$ and $\mat T_i=\widetilde{\mat S}_i\widetilde{\mat S}_i^H$. Since $\mat A^H\mat A=\mat I_R$, we have $\mat P_{\mat A}=(\mat A^H\mat C\mat A)^{-1}$. Removing the constant term and positive scale factor, \eqref{eq:fixed_target_problem} is equivalent to
\begin{equation}
\max_{\mat A^H\mat A=\mat I_R}
\operatorname{tr}\!\left[
(\mat A^H\mat C\mat A)^{-1}\mat A^H\mat T_i\mat A
\right].
\label{eq:fixed_target_trace_problem}
\end{equation}
Introduce $\mat Q=\mat C^{1/2}\mat A
(\mat A^H\mat C\mat A)^{-1/2}$.
Then $\mat Q^H\mat Q=\mat I_R$, and $\mat A$ can be recovered as $\mat A=\mat C^{-1/2}\mat Q(\mat Q^H\mat C^{-1}\mat Q)^{-1/2}$. Thus, the original problem can be equivalently reformulated in terms of $\mat Q$. Since $\mat C^{-1/2}=\mat L_H\mat B_\xi\mat L_H^H$, cyclic invariance of the trace and \eqref{eq:fixed_target_weighted_matrix} yield
\begin{equation}
\max_{\mat Q^H\mat Q=\mat I_R}
\operatorname{tr}\!\left[
\mat Q^H\mat L_H\mat M_i\mat M_i^H\mat L_H^H\mat Q
\right].
\label{eq:fixed_target_ky_fan_problem}
\end{equation}
By the Ky Fan maximum principle \cite[Th.~1]{fan1949weyl}, the maximum is attained by $R$ orthonormal eigenvectors associated with the largest eigenvalues of the Hermitian matrix. Thus, we may take $\mat Q^\star=\mat L_H\mat E_{i,R}$. Substituting into the inverse transformation yields \eqref{eq:exact_subspace_update}, proving the global optimality of $\mat A^\star$.
This completes the proof.
}

\section{Proof of Proposition~\ref{prop:stability}}
\label{app:proof_stability}
Let $\mat G_\ell=\mat H\bar{\mat F}_\ell\bar{\mat F}_\ell^H\mat H^H$ and $\mat W_\ell=\xi(\mat G_\ell+\xi\mat I_K)^{-1}$. {Apply $\mat A^{-1}-\mat B^{-1}=\mat A^{-1}(\mat B-\mat A)\mat B^{-1}$.}
{
Since $\mat G_\ell\succeq\mat0$, $\|(\mat G_\ell+\xi\mat I_K)^{-1}\|_2\le1/\xi$. Define $c=\|\mat H\|_2^2\|\bar{\mat F}_2\bar{\mat F}_2^H-\bar{\mat F}_1\bar{\mat F}_1^H\|_2/(KT\xi)$. Submultiplicativity gives $\|\mat W_1-\mat W_2\|_2\le KTc$. For the fixed-target cost $q_\ell(\widetilde{\mat S})=\operatorname{tr}(\widetilde{\mat S}^H\mat W_\ell\widetilde{\mat S})/(KT)$,
\begin{align}
|q_1(\widetilde{\mat S})-q_2(\widetilde{\mat S})|
&\le \frac{\|\widetilde{\mat S}\|_F^2}{KT}\|\mat W_1-\mat W_2\|_2\le c\|\widetilde{\mat S}\|_F^2.
\end{align}
Writing $f_\ell=f(\bar{\mat F}_\ell)=\min_{\widetilde{\mat S}\in\mathcal D(\mat S)}q_\ell(\widetilde{\mat S})$, feasibility of $\widetilde{\mat S}_2^\star$ for both problems gives
$f_1-f_2\le q_1(\widetilde{\mat S}_2^\star)-q_2(\widetilde{\mat S}_2^\star)\le c\|\widetilde{\mat S}_2^\star\|_F^2$.
Interchanging the indices gives $f_2-f_1\le c\|\widetilde{\mat S}_1^\star\|_F^2$. Combining the two inequalities yields $|f_1-f_2|\le cE_{\widetilde S}$, proving \eqref{eq:subspace_stability}.
}
{This completes the proof.}

\bibliographystyle{IEEEtran}
\bibliography{references}

\end{document}